\documentclass{IEEEtran}
\usepackage{amsmath}
\usepackage{amssymb}
\usepackage{graphicx}
\usepackage{xcolor}
\graphicspath{{./Figures/}} 
\DeclareGraphicsExtensions{.pdf,.jpg,.png,.jpeg}
\usepackage{float}
\usepackage{algorithm}
\usepackage{algorithmic}

\usepackage[
style=ieee,doi=false,isbn=false,url=false,eprint=false,maxnames=4
]{biblatex}
\begin{document}
	
\title{{\normalsize Author preprint July 2026}\\
Extracting Resilience Events from Utility Outage Data Based on Overlapping Times and Locations
}

\author{ Arslan Ahmad, Ian Dobson \\ Iowa State University
\thanks{A. Ahmad and I.~Dobson are with the Department of Electrical and Computer Engineering, Iowa State University, Ames Iowa USA; email: dobson@iastate.edu. 
Support from NSF grants 2153163 and 2429602, Argonne National Laboratory, and PSerc project S110 is gratefully acknowledged.}}

\maketitle	

\begin{abstract}
To study power system resilience with real data, it is necessary to group individual power outages recorded by utilities into events in which outages cluster and overlap due to extreme weather.
We show how to automatically group utility outage data into resilience events based on their time and location. 
Each outage is represented as a cylinder in three-dimensional space, with a disk centered at the outage location in the geographic plane and a vertical extent corresponding to a limited outage duration, so that two outages overlap in time and space if their cylinders intersect.
The grouping algorithm can be implemented as a graph whose nodes are the outages and whose edges represent the overlaps of outages in time and space, so that events are the connected components of the graph.
Extending time-based grouping to both time and location is particularly useful when extracting events from outage data collected across a wide area, as it prevents unrelated outages from being incorrectly merged into anomalous events solely due to temporal overlap. 
We propose a metric to tune the parameters of the grouping algorithm to minimize anomalous events.
The grouping of outages into events works with both detailed utility outage data and web-scraped EAGLE-I outage data.
Results are validated against NOAA storm event records and DOE OE-417 reports, and the automatically extracted events from utility data closely match documented major weather events.
\end{abstract}

\section{Introduction}
Power systems face increasing stress from extreme weather events, including severe storms, hurricanes, nor'easters, floods, wildfires, earthquakes, ice storms, derechos, geomagnetic storms, extreme temperatures, cyber attacks, physical attacks, cascading failures, and compound hazards \cite{doe417, climatecentral22, AnkitSSR22, LarsenENERGY16}. These events cause widespread outages that challenge utilities' ability to maintain reliable service and recover quickly. The study of power system resilience addresses this challenge by quantifying how well a system withstands and recovers from such large-scale disruptions. A key characteristic of such disruptions is that they cause multiple outages to occur within a relatively short period. That is, the outages bunch up and overlap \cite{WardCC13, MorrisPMAPS16}. Utilities typically have enough resources to restore power promptly following a few routine outages. However, during large events when multiple outages occur in a short time, it becomes challenging to restore them simultaneously, making the restoration process complicated and dependent on multiple factors. It is therefore pertinent that, instead of analyzing individual outages in large events, resilience analysis treats them as a single event to capture their combined impacts.

An important distinction exists between an individual outage and a resilience event. A single outage record describes the loss of a power system component at a specific time and location. It carries no information about whether that outage is part of a large disturbance or an isolated, routine failure. Resilience events, on the other hand, are groups of multiple outages arising from a common cause and occurring within a bounded window of time and space. 
Resilience metrics such as nadir, event frequency, total energy unserved, and restoration rate are inherently event-level quantities defined over a group of outages \cite{DobsonPS23, henryRESS12,NanRESS17,Pantelibook26,PanteliProcIEEE17,StankovicPS23}. 

It is therefore foundational for any \textit{data-driven} resilience analysis\footnote{In contrast, {\sl simulations} of resilience naturally produce outages that are already grouped into events.} to first group individual outage records into resilience events.   
Once events are identified, it becomes straightforward to compute a full range of event-level metrics and statistics that describe system resilience \cite{PapicAS20,CarringtonPS21,EkishevaPMAPS22,AbdelmalakIACCESS23,DobsonPS24,AhmadPS24,,LeeIACCESS24,ahmadArxiv26a}. Since utility outage datasets are extensive, often containing tens of thousands of records spanning multiple years, any systematic analysis requires automatic grouping. That is, it is impractical to manually group outages into events beyond some limited number of selected cases.

Using outage timing to group the outages into events has been shown to be quite effective in existing literature \cite{CarringtonPS21,EkishevaPMAPS22,AbdelmalakIACCESS23,DobsonPS24,AhmadPS24}. In particular, an automatic algorithm based on timing can identify large events that correspond well with extreme weather. 
However, events identified only by timing may include some outages that are geographically far enough from the main group to be clearly unrelated to it, and it is very desirable to exclude these outliers.
Therefore, in this work, we improve and extend grouping from time-based to both time- and location-based. 
The principle of overlapping for grouping outages into events is also clarified. 
Discriminating events with both time and location is particularly useful when processing utility data over a wide region.

In summary, this work develops novel methods that use \textit{only utility outage data} to \textit{automatically} group outages into events using \textit{outage location as well as time}, enabling further event-based quantitative resilience analysis. 
We test the proposed method using publicly available detailed outage data from investor-owned utilities in Massachusetts \cite{MassachusettsData}. Results are validated against NOAA (National Oceanic and Atmospheric Administration) storm event records \cite{noaaStormsDB} and DOE OE-417 reports \cite{doe417}. The top ten automatically extracted events closely match documented major weather events\footnote{We have not established that wildfires are addressed by our methods since they move in time and location differently.}. 
We also test the proposed method using the web-scraped EAGLE-I (Environment for Analysis of Geo-Located Energy Information) outage data \cite{eagleiWebsite} that describes customers outaged at all counties across the USA. Grouping the EAGLE-I data into events extends county-level analyses to capture the spatial structure of widespread events.

\section{Review of Existing Methods}
Identifying resilience events from outage data is a prerequisite for quantitative resilience analysis, yet the problem remains only partially standardized. The existing standards provide guidance on collecting, categorizing, and reporting distribution interruption data \cite{IEEEstd1782}, but they do not prescribe a general method for grouping individual outages into resilience events. Similarly, utility resilience planning practices recognize the importance of characterizing major events, but event definitions often remain vague, inconsistent, or tied to particular hazards \cite{keenNREL24, keenNREL24b, keenNREL24c}. This lack of a common event-definition procedure is consequential because many resilience metrics, including event frequency, event magnitude, nadir, duration, restoration rate, and customer-hours interrupted, are defined at the event level rather than at the level of individual outage records \cite{DobsonPS23, henryRESS12,NanRESS17,Pantelibook26,PanteliProcIEEE17,StankovicPS23}. Therefore, before outage data can be used for systematic resilience assessment, individual outages must be grouped into meaningful events.

A common approach in the literature is to define outage events using only temporal information. 
Carrington et al. \cite{CarringtonPS21} extract resilience metrics from distribution outage data by using the outage and restore processes and performance curve associated with an event. 
The performance curve tracks the number of unrestored outages.
In this framework, an event is defined by identifying the start and end of an event when the performance curve passes and returns to a threshold number of outages. 
(For a threshold value of zero outages, as assumed in \cite{CarringtonPS21}, this accumulation of outages until all are restored is equivalent to the events based on time in subsection \ref{sec:eventstime} with infinite $t_{\max}$, i.e., with no time limitation applied.)
This approach is well-suited to detailed outage records that include outage start and restoration times, and it provides a direct connection between event extraction and resilience metrics based on performance curves. 
Abdelmalak et al. \cite{AbdelmalakIACCESS23, AbdelmalakRW22} apply a related principle to EAGLE-I outage data, which is aggregated at the county level, by identifying events using thresholds on the fraction of customers interrupted (specifically, 5\% of a county's customers). 
Similar thresholds based on interrupted customers are also used in \cite{ShuaiEST25, RahmanERC25, StanishevskaArxiv25} for event identification and to distinguish significant events from routine background interruptions. 
These methods are attractive because they are simple, automated, and require only outage and restore or performance curve time series. They are also consistent with the intuition that resilience events correspond to periods during which outages cluster in time due to a common external driver.

Time-based grouping has also been used in transmission resilience studies. 
Papic et al. \cite{PapicAS20} group transmission outages into events by requiring outages to overlap in time and by imposing a short maximum separation (2 minutes) between the starting times of successive outages. 
Ekisheva et al. \cite{EkishevaPMAPS22} use a related time-based grouping procedure with a maximum time parameter (similar to the method discussed in subsection \ref{sec:eventstime} with $t_{\max}$=~1 hour), but also including successive outages starting within 5 minutes and excluding some repeats of momentary outages.
This approach has been adopted in NERC State of Reliability reporting to identify large transmission resilience events \cite{nercSOR23}. These studies demonstrate that temporal clustering is a practically useful basis for event extraction, particularly when the study region is limited or when the baseline rate outages is low enough that temporally overlapping outages are likely to be related.

The time-only grouping, however, has a critical limitation: outages that are geographically remote from the main disturbance can be incorrectly merged into the same event solely because they overlap in time.
As the spatial extent of the dataset grows, the probability that unrelated outages overlap in time also increases. A storm affecting one part of a service territory, a separate disturbance elsewhere, and isolated routine outages can therefore be merged into a single event simply because their outage intervals overlap in time. This over-aggregation can inflate event size, restoration duration, geographic extent, and customer impact, and can bias downstream resilience analyses. The problem is especially acute for statewide, regional, or entire interconnection datasets, where weather systems may affect only part of the study area while unrelated outages continue to occur elsewhere. Thus, while temporal overlap is necessary for many event definitions, it is not always sufficient to establish that outages belong to the same resilience event.
All of these time-based grouping variations can be extended to location-based grouping as well, using the cylinder concept from subsection~\ref{sec:eventstimelocation}.

A second class of methods addresses this limitation by using weather information to define the event first and then associating outages with the weather-defined event. In these weather-first approaches, the analyst identifies a severe weather episode, determines its time window and geographic footprint, and then collects outages that occur within those boundaries \cite{DunnNHR18, LeeIACCESS24, leeIEEEaccess25, DonaldsonCIRED26}. This strategy is appropriate for studying well-defined hazards such as named hurricanes, major winter storms, or high-wind events, and it can directly connect outage impacts to meteorological drivers. Lee et al. \cite{LeeIACCESS24}, for example, use weather advisories and warnings to identify extreme outage events, and define extreme events by grouping all outages occurring 3 hours before to 48 hours after the end of the weather event (using the NWS Valid Time Extent Code dataset and the EAGLE-I dataset). Such approaches are valuable when the goal is to study the impact of specific types of weather events.

Weather-first methods nevertheless introduce additional challenges. They require external meteorological data that must be available at appropriate spatial and temporal resolutions and aligned with outage records. 
Defining the start time, end time, and spatial footprint of a weather event can be ambiguous, especially for compound, slow-moving, or transitioning storms. 
Moreover, weather-first procedures may focus attention on large events with clear meteorological signatures, while missing moderate but operationally significant disturbances that are clearly defined in outage data but not easily captured in weather data because the boundaries in time and space of the moderately bad weather are hard to determine. 
These limitations motivate outage-data-driven approaches that can systematically identify events without relying on the weather data. 
The merits of the different weather and outage based approaches to defining events and the wider policy uses of these events are discussed and illustrated in more detail in \cite{DonaldsonCIRED26}.

Several studies provide evidence that outage and fault data contain spatial as well as temporal structure. Morris et al. \cite{MorrisPMAPS16} analyze fault events on the Great Britain transmission network and show that weather-related faults exhibit spatial clustering and temporal clustering. Their work supports the premise that outage data itself encodes spatial and temporal structure that can be extracted directly, without requiring external weather datasets. However, their analysis is primarily descriptive and does not provide a general automated procedure for grouping outages into spatiotemporal resilience events. More recently, Wang et al. \cite{wangSR26} combine temporal overlap with spatial aggregation using Voronoi polygons associated with weather stations. This recognizes the importance of both time and location, but the spatial aggregation remains tied to weather-station regions rather than being formulated as a general event-extraction rule based directly on outage locations.

The method proposed in this work addresses limitations in the existing research.
The proposed method uses only outage data and groups outages using both temporal overlap and spatial proximity. 
This formulation generalizes existing temporal-overlap methods: if the spatial condition is removed or made nonbinding, the method reduces to time-based grouping; when spatial proximity is enforced, geographically unrelated outages are prevented from being merged into the same event.

The contribution is also distinct from weather-first event identification. Rather than using weather data to define events and then searching for associated outages, the proposed method extracts events directly from the outage records. Weather information can then be used afterward for validation or interpretation. This distinction is important because it allows the method to be applied consistently across different outage datasets, weather types, and event severities. It also enables event extraction when detailed weather-event boundaries are unavailable or difficult to specify.

The method is designed to apply to both detailed utility outage records and county-level web-scraped \text{EAGLE-I} outage data. For detailed utility data, spatial proximity is measured using outage locations, and temporal proximity is based on overlapping outage intervals with a maximum duration parameter to prevent very long individual outages from unrealistically linking otherwise separate events. For EAGLE-I data, which report interrupted customers at county-level time samples rather than individual outage records, the same conceptual approach is adapted by first defining county-level outage events and then grouping neighboring county events that overlap in time. This provides a unified framework for event extraction from both high-resolution utility records and web-scraped  aggregated outage data.

The proposed method, therefore, addresses a specific gap in the existing literature. Prior time-based methods provide useful automated event extraction, but can over-aggregate events over wide areas. Weather-first methods provide strong physical interpretation but require external weather data and event-boundary assumptions. Descriptive spatiotemporal studies demonstrate the presence of clustering but do not by themselves supply a general event-extraction algorithm. The method proposed here combines time and location in a direct, outage-data-driven grouping rule, provides a graph-theoretic implementation through connected components, and includes threshold-tuning procedures to minimize anomalous groupings. This places the current paper in the research space as a general, practical method for automatically extracting resilience events from outage data, suitable for subsequent quantitative resilience analysis.

The remainder of this paper is organized as follows: Section~\ref{sec:data} introduces the data used in this work, 
Section~\ref{sec:eventsFromDetailedData} defines events based on time, location, and the combination of both in the detailed outage data, 
Section~\ref{sec:eventsFromEagleiData} explains how the proposed methods can be adapted for EAGLE-I data. Section~\ref{sec:tuning} discusses the tuning and sensitivities of the thresholds used in the event grouping, Section~\ref{sec:results} presents results from the detailed outage data and aggregated EAGLE-I data, and Section~\ref{sec:conclusion} concludes.

\section{Data}\label{sec:data}
The proposed methods in this work are tested using actual power outage data from two sources: 1) Detailed power outage data from all the investor-owned utilities in the state of Massachusetts \cite{MassachusettsData} from 2013 to 2023, and 2) Web-scraped outage data from EAGLE-I \cite{eagleiWebsite} from 2014 to 2025.
The detailed outage data provides information about individual outages, including their exact start and end times, the number of customers affected, and their locations. 
On the other hand, the EAGLE-I data has the advantage of recording outages from almost all counties in the US; however, it only samples interrupted customers every 15 minutes and is aggregated at the county level, so we do not have information about exactly where the interrupted customers are located within a county. 
Given the limitations in the temporal and spatial coverage of the EAGLE-I data, it is properly processed first to ensure consistent data before use in this analysis.

\section{Events in Detailed Outage Data}\label{sec:eventsFromDetailedData}

\begin{figure*}
    \centering
    \includegraphics[width=\textwidth]{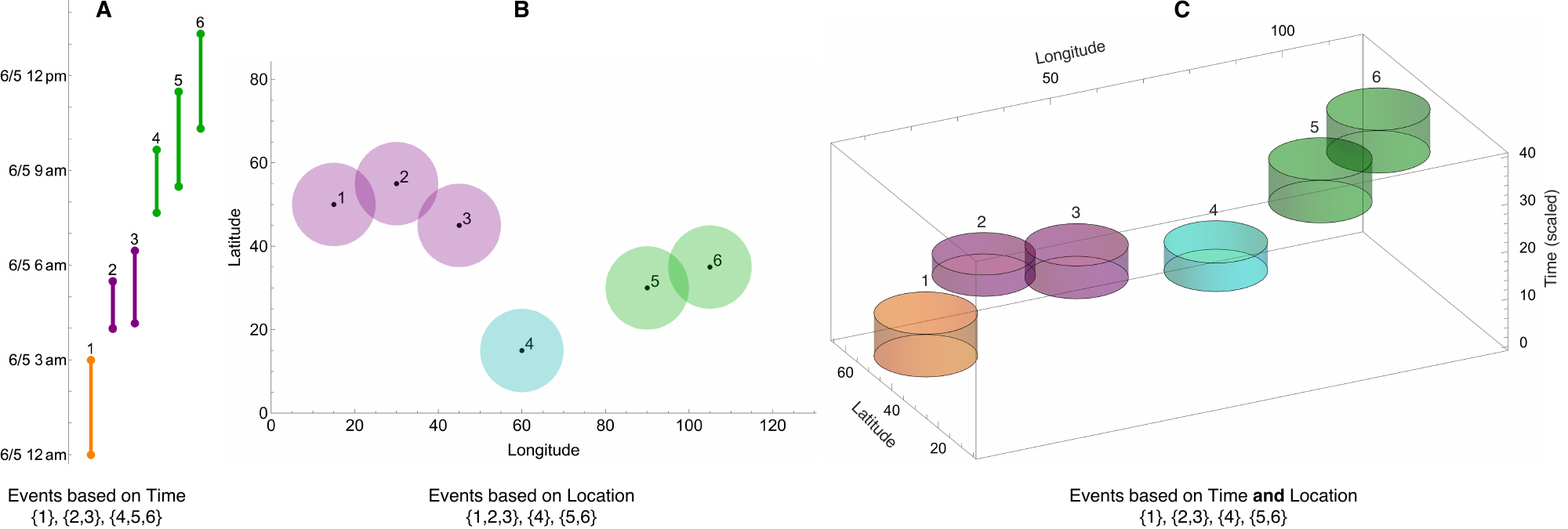}
    \caption{Example of six outages resulting in different events in time, location, and time and location together. The distance threshold is $d=20$ km.}
    \label{fig:eventsPlot}
\end{figure*}

\subsection{Events based on time}
\label{sec:eventstime}

Each outage occurs over a time interval $[o,r]$ where $o$ is the outage start time and $r$ is the outage restore time. Two outages overlap in time if their time intervals overlap; that is, if one of the outages starts before the other outage is restored. Then events are a maximal group of outages that are connected together by overlapping in time.
That is, two outages are in the same event if they are connected by a series of overlapping outages.
For example, in Fig.~\ref{fig:eventsPlot}A, outage 1 does not overlap with any other outages, so it forms an event with only one outage. Outages 2 and 3 overlap but do not overlap with any other outages so they form an event with two outages. Outages 4 and 5 overlap, and outages 5 and 6 overlap, due to which outages 4, 5, 6 are in the same event (in this case, outages 4 and 6 need not overlap).

\begin{figure}
    \centering
    \includegraphics[width=1.0\linewidth]{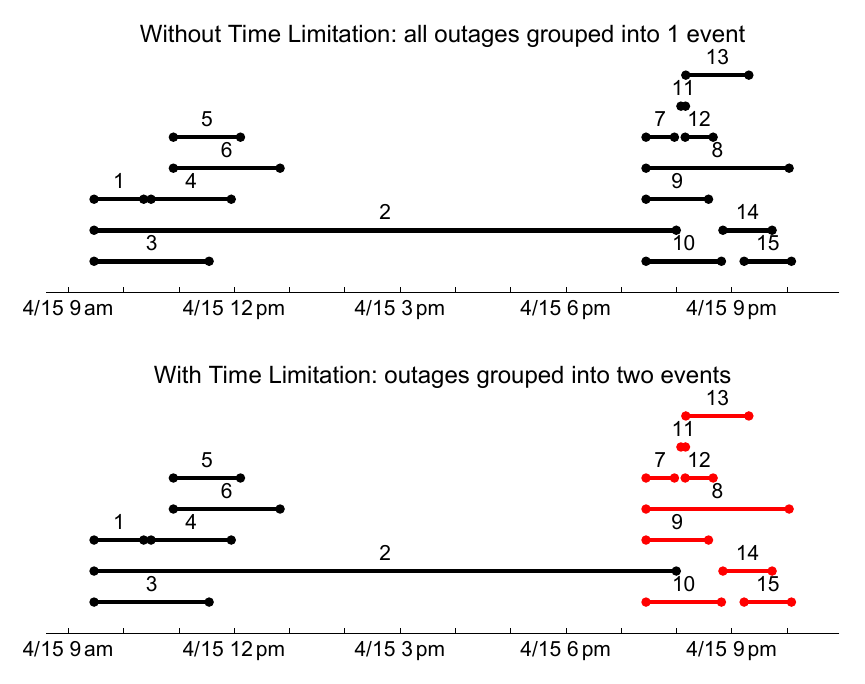}
    \caption{
    The top plot shows an unrealistic long event where the abnormally long duration of outage 2 causes all the outages to be incorrectly grouped into a single event. The bottom plot shows how a time limitation of $t_{\rm max}=3 $h fixes this problem and correctly groups the outages into two distinct events (shown in black and red color), while retaining the actual duration of outage~2.}
    \label{fig:veryLongEvent}
\end{figure}

The definition of events in time is often slightly more elaborate than explained so far, in order to limit the effect of outages with very long restoration times.
Instead of associating each outage with the time interval $[o,r]$, each outage is associated with the time interval $[o,\min\{r,o+t_{\rm max}\}]$, limiting the time interval to a maximum duration of $t_{\rm max}$ hours. 
These limited time intervals are then used to define the events. 
This time restriction enables grouping outages into events based on temporal overlap, ensuring that a few very long outages do not result in overly long, unrealistic events; an example of such an unrealistic long event is given in Fig~\ref{fig:veryLongEvent}.
It is imperative to note that while the time limitation is used to define the overlapping outages in an event, the full outage duration is preserved when the event is analyzed.

\begin{figure}
    \centering
    \includegraphics[width=1.0\linewidth]{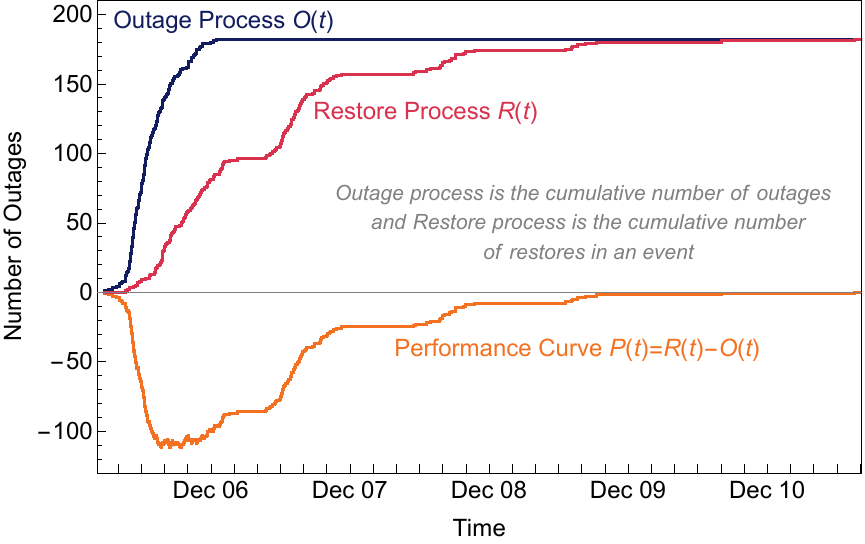}
    \caption{Performance curve $P(t)$, outage process $O(t)$, and restore process $R(t)$ of an event with 184 outages.}
    \label{fig:eventProcesses}
\end{figure}

Another way to characterize events is through the performance curve $P(t)$, which tracks the negative of the cumulative number of unrestored outages \cite{CarringtonPS21}.
An event begins with its first outage and is followed by a series of outages with overlapping time intervals until all the outages are restored. 
Therefore, $P(t)$ starts at zero outages, is negative during the event, and first returns to zero at the end of the event.
Indeed, the number of unrestored outages $-P(t)$ is the number of overlaps at time $t$.
An example performance curve of an event is shown in Fig.~\ref{fig:eventProcesses}.

\subsection{Events based on location}
\label{sec:eventslocation}
In addition to the temporal information, we also have spatial information for each outage in the outage data, usually in the form of latitude and longitude coordinates.
Therefore we know when an outage started, when it was restored, and where it occurred.
Using this information, outages can be grouped together into events based on the distances between them.
Two outages are close if their distance is less than or equal to a threshold distance $d$. Equivalently, associate each outage with a hypothetical disk of radius $d/2$ centered at the outage location so that the outages are close if their disks overlap.
Then events are a maximal group of outages that are connected together by their disks overlapping.
That is, two outages are in the same event if they are connected by a series of overlapping outage disks.
In Fig.~\ref{fig:eventsPlot}B, the 4 events based on location are \{outages 1,2,3\}, \{outage 4\}, and \{outages 5,6\}.

It should be noted that the proposed location-based grouping has a ``chaining" effect. We explain this with Fig~\ref{fig:eventsPlot}: outage 3 is not within $d$ of outage 1, yet they are considered part of the same event because outage 2 connects them. Therefore, the average size of location-based events often exceeds $d$. 
If $d$ is set too large, this chaining can lead to over-aggregation in which spatially unrelated outages can become linked through a chain of intermediate outages. This motivates the combined time-and-location method described in Section \ref{sec:eventstimelocation} and the threshold tuning procedure in Section \ref{sec:tuning}.

The grouping based on location works with several specifications of distance. 
If location is given in the data as latitude and longitude, Euclidean or haversine distance can be used. 

\subsection{Events based on time and location}
\label{sec:eventstimelocation}
\begin{figure}[htb]
    \centering
    \includegraphics[width=1.0\linewidth]{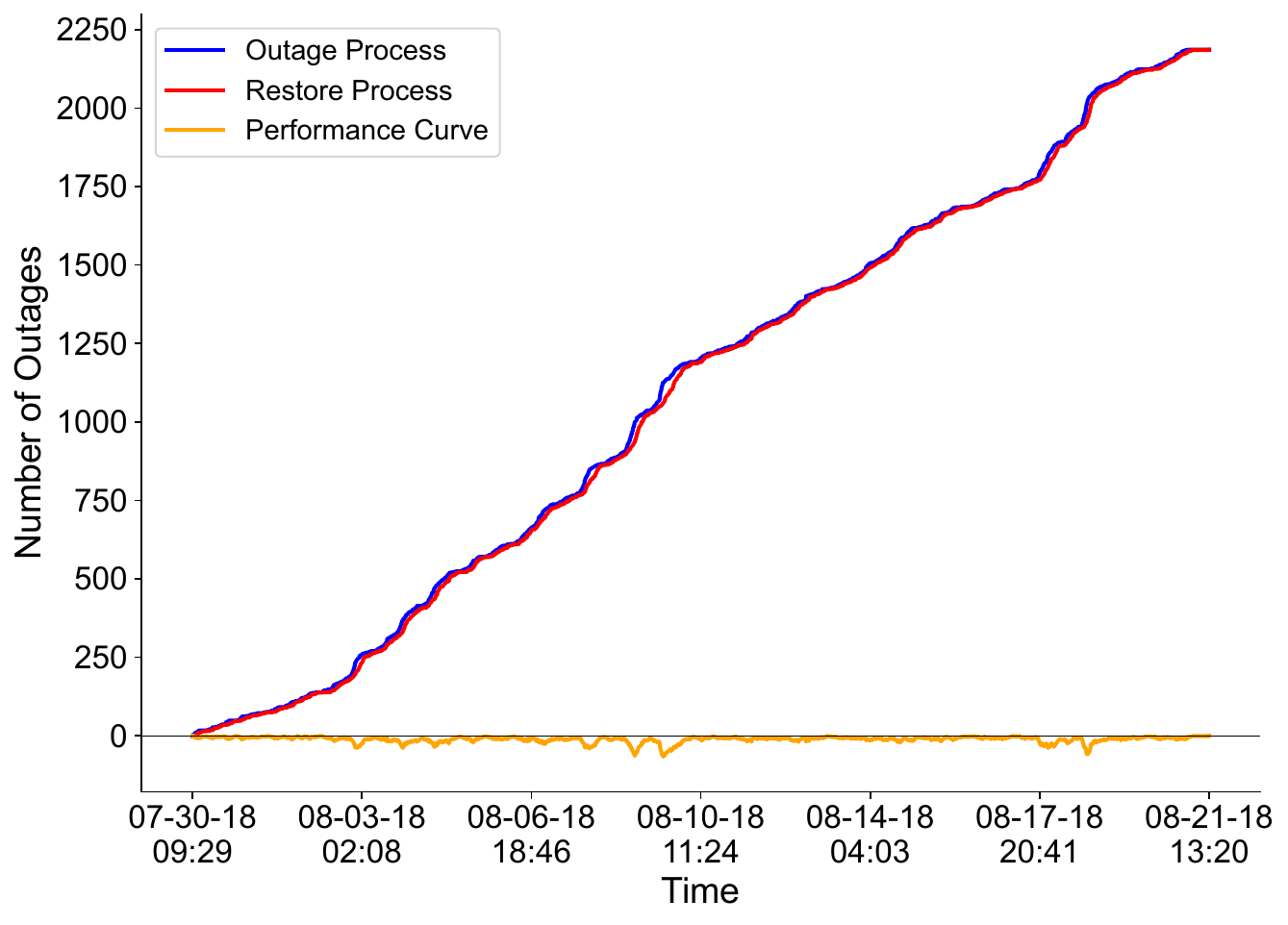}
    \caption{Example of an anomalous event with a very long duration ($\approx$22 days) and a small area under the performance curve. These types of events, referred to as `snake events', are generated in the data when outages from a large geographic area are grouped together without integrating outage location information into the grouping algorithm.}
    \label{fig:snakeEvent}
\end{figure}

When analyzing outage data from a large geographic area, time-based grouping can combine unrelated outages that are very far from the main event. When analyzing such large areas, the probability of an outage within a given time increases. As a result, a time-based grouping algorithm continues to add subsequent outages to the same event, and the event keeps growing due to over‑aggregation. 
An example of such an event is shown in Fig.~\ref{fig:snakeEvent}. This event has a large number of outages and its performance curve is consistently small, with no dominant hump.
Particularly, the performance curve size is very small relative  to the number of outages.
The event does not resemble a typical outage event (such as the one shown in Fig.~\ref{fig:eventProcesses}), and appears visually like a snake, which is why we refer to this type of anomalous event as a `snake event'.

To overcome this problem, we account for both time and location by defining an event as a group of outages that both \textit{overlap in time} and are \textit{close in distance}.
We consider the location and time information of outages as a 3-dimensional space, where the $x$ and $y$ axes represent location, and the vertical $z$ axis represents time. In the 3-dimensional space, each outage is associated with a cylinder with base radius $d/2$ and a vertical time dimension given by the limited outage duration $[o,\min\{r,o+t_{\rm max}\}]$ as shown in Fig.~\ref{fig:eventsPlot}C.
Two outages are close in time and location if their cylinders overlap in 3 dimensions; that is, both their limited durations and disks overlap.
Then, as before, an event is a maximal group of outages that are connected together by overlapping.
In Fig.~\ref{fig:eventsPlot}C, the 4 events based on time and location are \{outage 1\}, \{outages 2,3\}, \{outage 4\}, and \{outages 5,6\}.

\subsubsection{Graph implementation}
A straightforward way to implement the time and location based event extraction from data is to create an undirected graph in which outages are represented as nodes, and outages that overlap in time and location are connected by edges. 
Let $X$ be the set of outages, where each outage is regarded as a node of an undirected graph $G=(X,E)$.
An edge connects any two nodes in $G$ if the corresponding two outages overlap in time and are close in location:
\begin{equation}
    E=\{(i,j): I_i \cap I_j \neq \varnothing ~{\rm and}~ d_{i,j} \leq d \}
\end{equation}
where
\begin{equation}
    I_i = [o_i,\min\{r_i,o_i+t_{\rm max}\}]
\end{equation}
Then events are the connected components or maximal connected subgraphs of $G$\footnote{There is also a concept of connected or path-connected components in topology, giving an elegant definition of events: Events are the outages in the connected components of the union of all the outage cylinders.}.
The graph formulation lends itself to optimization for larger datasets using standard techniques such as interval trees, sweep line, k-d trees, and grid hashing. 

The parameters $t_{max}$ and $d$ constrain only pairwise links between outages; because events are connected components, the total duration and geographic span of an event may exceed these thresholds through chaining. Indeed, the chaining allows the storm outages formed by a weather front moving across a large area to be grouped together.

\section{Events in EAGLE-I Outage Data}\label{sec:eventsFromEagleiData}
The concepts of events based on time, location, and time-and-location apply to EAGLE-I data as well, though the processing has some differences due to how the EAGLE-I data is recorded. 
The EAGLE-I data samples a performance curve and does not record individual outages and thus does not provide individual outage start and restoration times.
Therefore, the $t_{\rm max}$ threshold cannot be used with the EAGLE-I data, and we define events using the performance curves.

The \hbox{EAGLE-I} data specifies a performance curve $P^{\rm cust}(t)$ tracking the number of unrestored customers in a county at 15-minute samples of time $t_1,t_2,\cdots$. 
To define county events based on time in the \hbox{EAGLE-I} data for Massachusetts, we set a threshold of 30 customers\footnote{The threshold is needed to distinguish the start and end of events while small number of customers are persistently outaged throughout the county, or while the recorded data is spuriously stuck at a small number of customers out. This threshold can vary from one county/state to another, and can also be selected as a percentage of the total customers in that county/state, as done in \cite{AbdelmalakIACCESS23}. A value of 30 customers appeared to be working well for the states of Massachusetts and Iowa.}.
A county event starts when $P^{\rm cust}(t)$ exceeds 30 customers and ends just before $P^{\rm cust}(t)$ first drops below 30 customers. That is, an \hbox{EAGLE-I} county event occurs at a maximal series of consecutive samples $t_m, t_{m+1},...,t_{m+n}$ for which 
 $P^{\rm cust}(t_k)>30$ for $k=m,m\!+\!1,...,m+n$.
An increment in apparent outages occurs at $t_k$ when $P^{\rm cust}(t_k)>P^{\rm cust}(t_{k-1})$. 
The duration of the outage process of a county event is the time samples between and including the first outage increment at the start of the county event and the last outage increment in the county event. 
Then events in two counties overlap in time when their outage processes overlap.

In contrast to the detailed outage data, which has the exact location of each outage, the EAGLE-I data only has the county location. For location-based event grouping in the EAGLE-I data, we treat counties as nodes in a graph and connect any two counties with a sufficiently long common border via a graph edge. Then neighboring counties have a graph distance of 1, and the threshold $d=1$ regards a county and its neighbors as close.

For time-and-location-based grouping in the EAGLE-I data, we treat all county events whose outage processes overlap and are also in neighboring counties as part of the same event. 
Therefore, there is only one parameter, the distance $d$, for time-and-location-based grouping in the EAGLE-I data.
We set $d=1$, i.e., two outages are considered spatially close if they are in adjacent counties. Since outages with successive overlaps have a ``chaining" effect, counties that are not immediate neighbors are also grouped into the same event if they are connected by a series of overlapping outages. 

\section{Threshold Selection and Sensitivity}\label{sec:tuning}

The time and location thresholds $t_{max}$ and $d$ influence the total number of events, the size of events, as well as the shape and other characteristics of events.
In this section, we discuss how different thresholds affect events, how to tune them to achieve good events, which events are considered anomalous, and how sensitive the results are to these thresholds.
Tuning is important for extracting a very high proportion of credible events. Poorly tuned events can significantly skew the resilience metrics calculated from those events.

We optimize threshold tuning based on the following objective: ``Capture the \textit{complete outage process} of an event with a \textit{minimum number of irrelevant outages} that do not belong to that event." 
We propose a hump metric $M$ in Section \ref{sec:minimizing} to operationalize this objective.
Since events are primarily required for resilience analysis, we focus our threshold tuning on obtaining accurate medium-to-large events to achieve this objective; only events with at least 20 outages are evaluated during the tuning process.

Different threshold choices could either merge outages that should be separate (over‑aggregation) or split one event that should remain combined into several events (over‑fragmentation). In particular, very small thresholds in time and location lead to over-fragmentation, while very large thresholds lead to over-aggregation. Smaller thresholds yield more events, most of which are single-outage events, and the largest events are, on average, smaller (with fewer outages and shorter outage process durations). On the other hand, larger thresholds yield fewer events and fewer single-outage events because smaller events combine into larger ones, and the average size of the largest events is larger (with more outages and longer outage process durations). 

\begin{figure}[htb]
    \centering
    \includegraphics[width=1.0\linewidth]{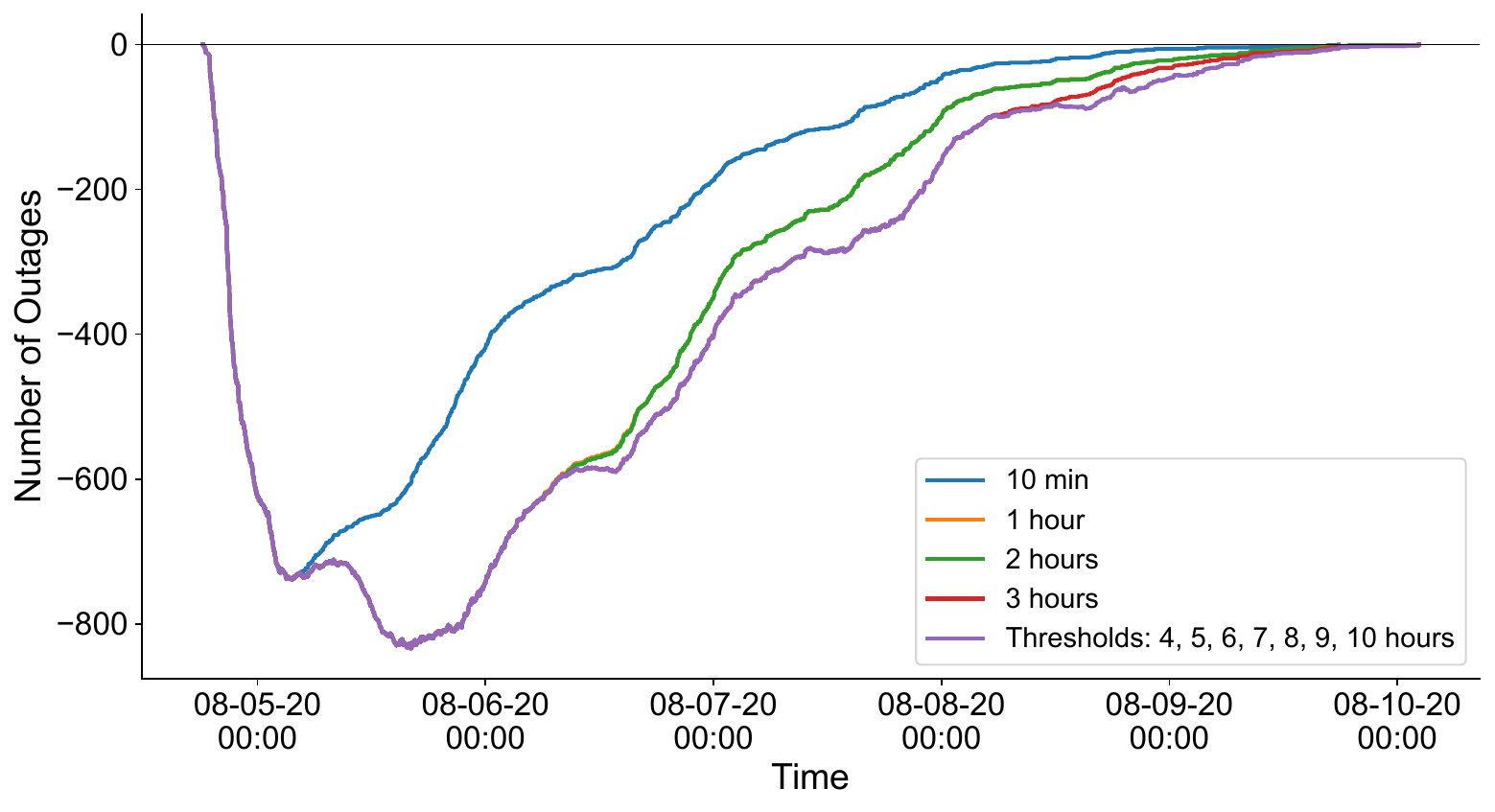}
    \caption{Performance curve of a large event generated with different time thresholds $t_{\rm max}$.}
    \label{fig:largest_event_performance_curves}
\end{figure}

The example in Fig.~\ref{fig:largest_event_performance_curves} provides a closer view of how changing thresholds affects a large event's performance curve. We can see the effect of over-fragmentation by comparing the curve at the $t_{\rm max}=$ 10 minute threshold with those at higher threshold values. 
It shows that the event continued, but the grouping algorithm stopped adding subsequent outages to the event because $t_{\rm max}=$ 10 minute was too low, leaving the event incompletely captured. 
On the other hand, when we compare the $t_{\rm max}=$ 2 hour curve with those at higher thresholds, we can see that the algorithm continued to add outages to the same event during its restoration phase. These additional outages may or may not be related to the same event. 
Lastly, we can see that the performance curve remains the same at $t_{\rm max}=$ 4 hour and above, indicating that increasing the thresholds beyond a certain point can have minimal effect on large events because there are very few or no additional outages close enough in time and location to be added to them.

\subsection{Minimizing anomalous events with the hump metric}
\label{sec:minimizing}
\begin{figure}[htb]
    \centering
    \includegraphics[width=1.0\linewidth]{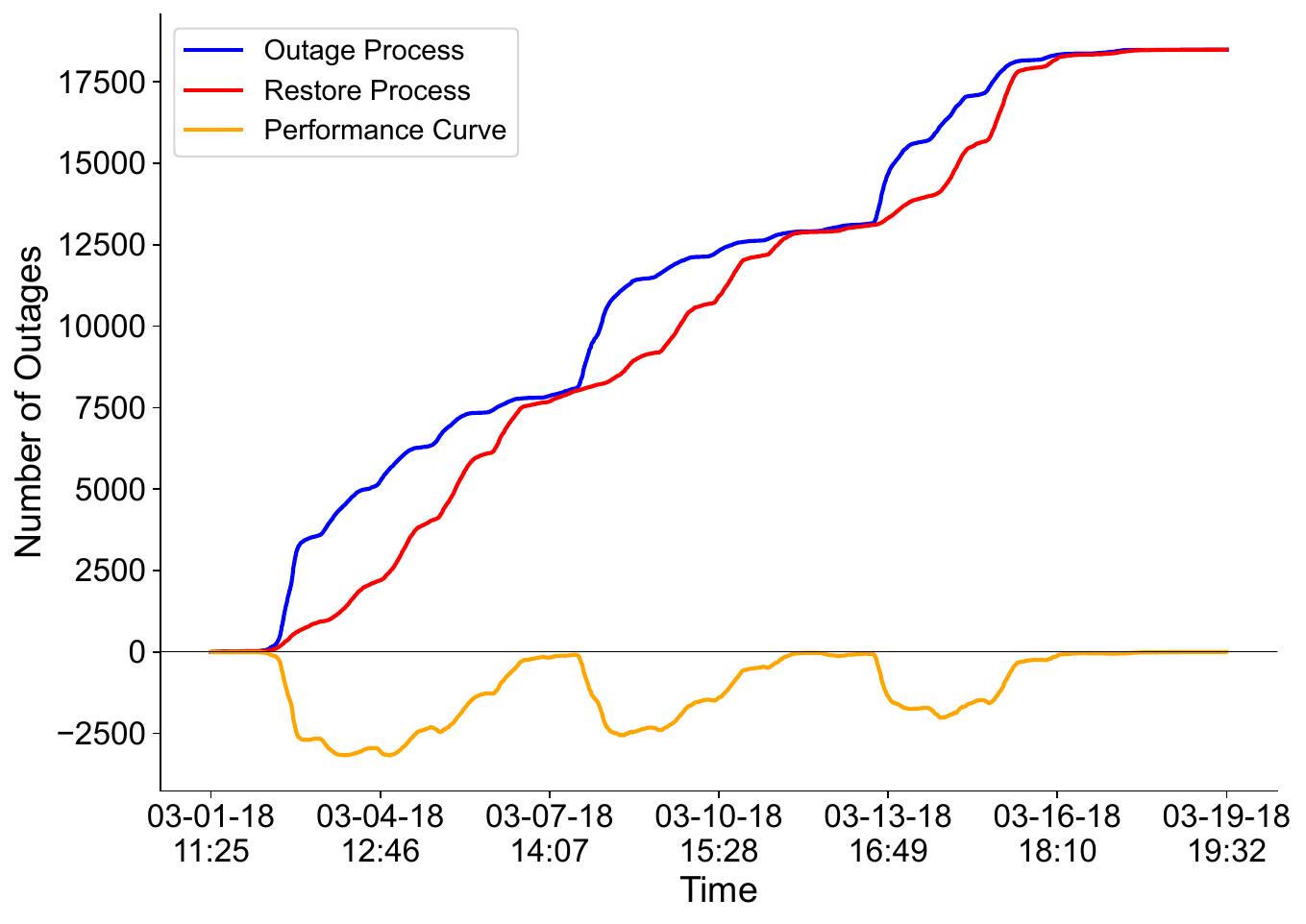}
    \caption{Example of an event with multiple humps showing three events incorrectly combined into one event.}
    \label{fig:multiple_hump_event}
\end{figure}

\begin{figure}[htb]
    \centering
    \includegraphics[width=1.0\linewidth]{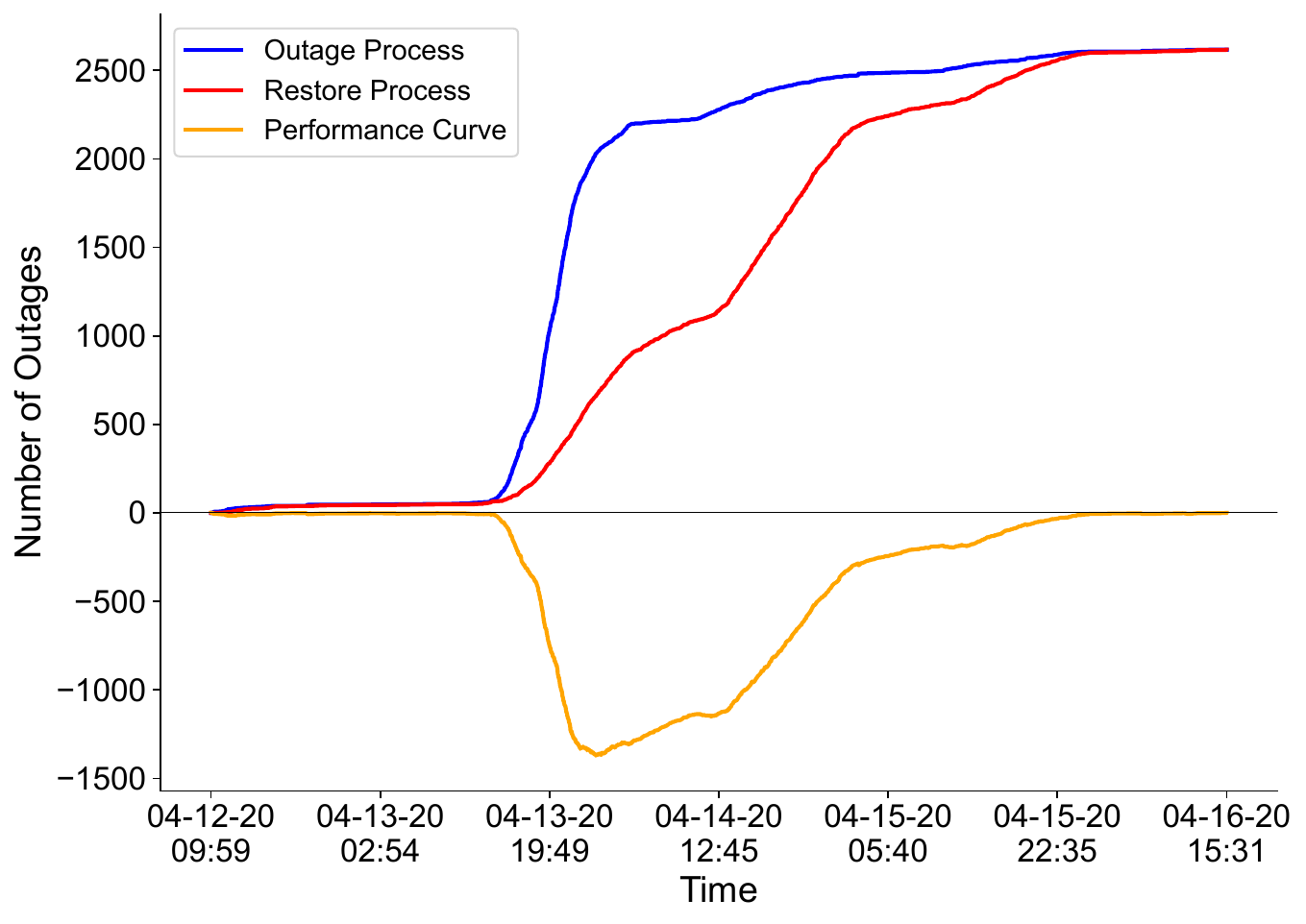}
    \caption{Example of an event with a long left `tail'.}
    \label{fig:left_tail_event}
\end{figure}

We want to select thresholds that give us medium-to-large events that minimize the number of anomalous events of the following types:
\begin{itemize}
    \item \textit{Snake Events}: As discussed in section \ref{sec:eventstimelocation}, when very large time or location thresholds are used, multiple outages that are not related to a single event are grouped into one event, which gives a very long, unrealistic event with the restore process following the outage process very closely and a consistently small performance curve, as shown in Fig.~\ref{fig:snakeEvent}. Snake events have a high ratio of event duration to the area under the performance curve. 
    \item \textit{Multi-hump Events}: When multiple events that happen in close succession are grouped together into one event, we see multiple `humps' in the performance curve, as shown in Fig.~\ref{fig:multiple_hump_event}. The algorithm needs to discriminate the multiple hump events that most likely should be separated into multiple events from those that should most likely remain combined into one event. 
    \item \textit{Left-tail Events}: If thresholds are not properly tuned, we get some events with long left `tail' at the start of their performance curves, as shown in Fig.~\ref{fig:left_tail_event}. These events are anomalous because they contain outages at the start that are likely not part of the event, thereby underestimating the start time and overestimating the duration of the event.
\end{itemize}

We propose a hump metric that can be used to tune the time and location thresholds by identifying and minimizing the number of the aforementioned anomalous events.
Let $O(t)$ and  $R(t)$ be the event outage process and restore process, respectively.
The outage process is the cumulative number of outages, and the restore process is the cumulative number of restores in an event, as shown in Fig.~\ref{fig:eventProcesses}.
We divide the outage and restore processes by their corresponding maximum values, which are the total number of outages and also the total number of restores, to normalize $O(t)$ and  $R(t)$ to the range $[0, 1]$. We denote the normalized outage and restore curves by $o(t)$ and  $r(t)$, respectively.
Then $p(t)=o(t)-r(t)$ is a scaled and positive version of the performance curve\footnote{Here we consider the performance curve to be positive, in contrast to the negative performance curves shown in all of the figures. This is done merely to ensure positive areas that simplify explanations.}.
We similarly normalize the time within an event to be in the range $[0, 1]$ and denote the normalized time as $\tau$:
\begin{equation}
    \tau = \frac{t-t_{start}}{t_{end}-t_{start}}
\end{equation}
Let $p_{\rm max} = {\rm max}~p(\tau)$ be the maximum of the scaled performance curve. 
We define a floor level $f=0.2\, p_{\rm max}$.
Let $A_1,A_2,A_3,...,A_m$ be the areas under the performance curve for all the time intervals in which $p(\tau)>f$ so that the performance curve is higher than the floor level.
More explicitly, if during the event the times with $p(\tau)>f$ form $m$ intervals and $[a_k,b_k]$ is the $k$th such interval,
then
\begin{align}
A_k=\int_{a_k}^{b_k}p(\tau)d\tau,\quad k=1,2,...,m.
\end{align}
$A_1,A_2,A_3,...,A_m$ are the areas under humps of the performance curve.
Let $K$ be the index of the maximum of $A_1,A_2,A_3,...,A_m$.
The hump metric is then defined as:
\begin{equation}
    M  = A_K - \sum_{\substack{k=1\\k\ne K}}^{m}A_k = 2\max_k A_k - \sum_{k=1}^{m}A_k 
    \label{eq:tuningMetric}
\end{equation}
The hump metric $M$ is a dimensionless quantity.
It is larger for events with one hump.
A lower value of  $M$ suggests anomalous events, as $M$ penalizes events with multiple humps, snake events with a consistently small performance curve throughout the event, and events with long tails. We regard events with $M<0.025$ as anomalous.
The value $M = 0.025$ is selected by inspecting the distribution of $M$ across all medium-to-large events: events identified visually as anomalous (snake events, multi-hump events, and long-tail events) consistently produced $M$ below this value, while credible single-hump events produced $M$ well above it.

\begin{figure}[htb]
    \centering
    \includegraphics[width=0.65\linewidth]{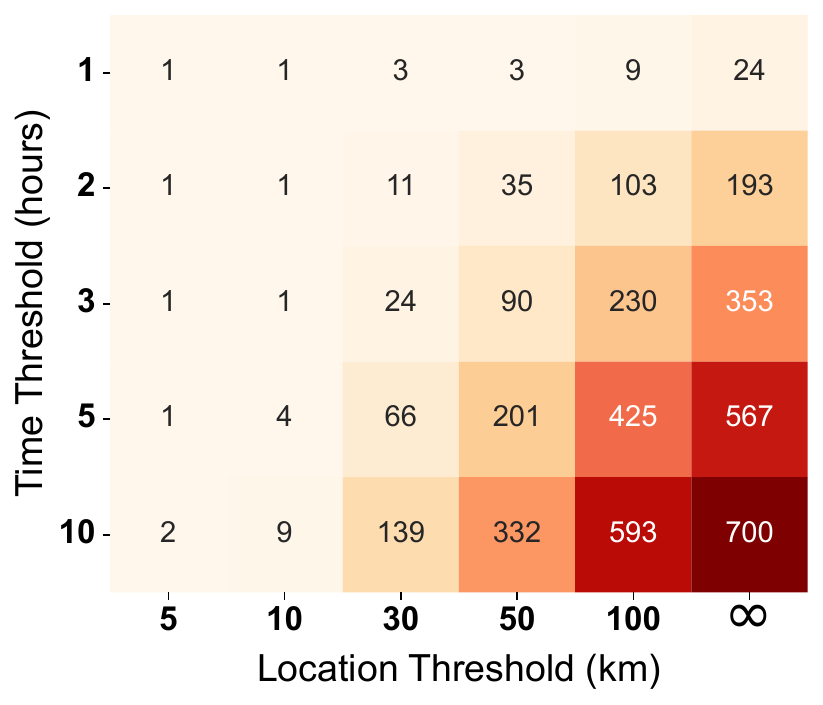}
    \caption{Heatmap of the number of anomalous events (those with $M<0.025$) corresponding to different combinations of the time threshold $t_{\rm max}$ and location threshold $d$. The rightmost column shows the numbers when only time-based grouping is used.}
    \label{fig:tuning_heatmap}
\end{figure}

$M$ is calculated for different combinations of the time and location thresholds, and the number of anomalous events is identified as shown in Fig.~\ref{fig:tuning_heatmap}.
To avoid over-fragmentation, we want to select the \textit{largest} time and location thresholds that yield the fewest anomalous events. For our cases, the values are $t_{max}=3$ hours and $d=10$ km $=6.2$ mi, which yields only one anomalous event.

For EAGLE-I data, there is only one parameter, $d$, to be tuned (with the county-adjacency rules and the county-event threshold fixed). 
After testing results across multiple counties and states, we found that the $d=1$ threshold works best for EAGLE-I;  selecting a higher threshold $d\ge2$ tends to group together unrelated outages.

\subsection{Threshold Sensitivity}
\begin{figure}[htb]
    \centering
    \includegraphics[width=1.0\linewidth]{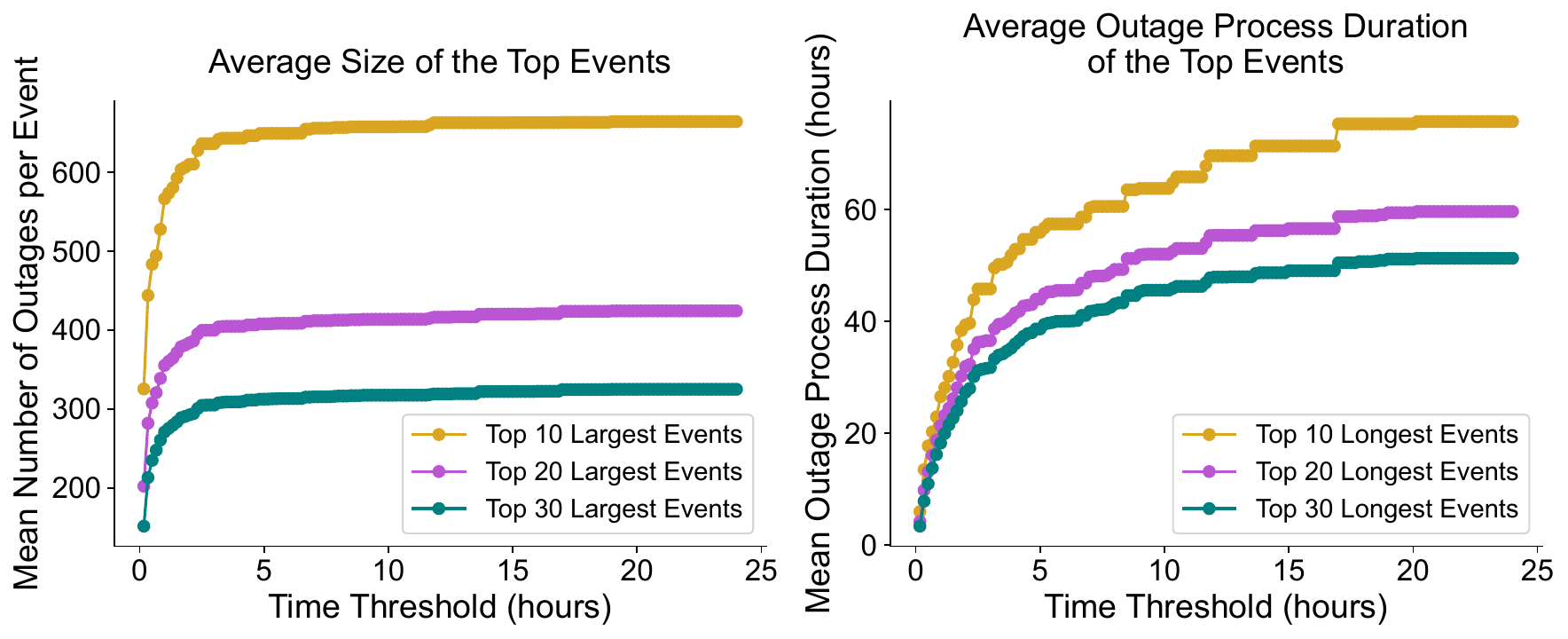}
    \caption{Sensitivity of the average size and average outage process duration of the top events for different values of the time threshold $t_{\rm max}$ from 10 minutes to 25 hours with 10-minute increments.}
    \label{fig:time_threshold_sensitivity}
\end{figure}

The number of events of different sizes, as well as their statistics and metrics, are sensitive to the choice of time and location thresholds. However, the sensitivity is high for lower threshold values, as shown in Fig.~\ref{fig:time_threshold_sensitivity}.
As we are interested in resilience events, we examine how the average size and average outage process duration of the top 10, 20, and 30 events change as the time threshold $t_{\rm max}$ changes for detailed outage data with infinite $d$.
At lower thresholds, there are more small events, increasing the likelihood that smaller events will combine to form larger ones. Whereas at higher thresholds, most outages associated with a single event are already grouped into large events, and we see only small changes as we continue to increase the threshold.
This effect is reflected as a steep slope of the curves in Fig.~\ref{fig:time_threshold_sensitivity} at the beginning, followed by an almost flat slope region showing lower sensitivity. 
The decrease in sensitivity as the time threshold increases supports the choice of the {\sl largest} time threshold that minimizes anomalous events in section~\ref{sec:minimizing}.

\subsection{Weather-informed event validation}
Resilience events involve a large number of outages which are often caused by extreme weather. Therefore, the spatial and temporal characteristics of different types of weather events (such as the typical duration of a tornado or the typical spatial spread of a hurricane) can provide approximate validation of the events obtained after threshold tuning, particularly when identifying  events associated with specific weather types.
Analysis of historical weather events  for the area under study reveals insights into the prevailing types of weather events and their typical spatio-temporal extents. Additionally, studies such as \cite{MorrisPMAPS16} give the average size of ice, snow, sleet, and blizzard clusters as $\approx 70$ miles, the average size of lightning clusters as $\approx 74$ miles, and the average size of wind, gale, and windborne objects clusters as $\approx 35$ miles.

The National Oceanic and Atmospheric Administration's (NOAA) Storm Events database \cite{noaaStormsDB} is a very helpful resource for finding information on different types of storms, including tornadoes, thunderstorms, floods, and snowstorms. Temporal and spatial information about specific types of events can be downloaded. The events can be independently extracted from outage data using the method discussed in this paper, and then compared with storm data to validate whether the chosen thresholds resulted in outage events that resemble the storm events. This method is used to validate the results discussed in Section \ref{sec:results}.

\section{Results}\label{sec:results}
\subsection{Detailed outage data results}
The time-and-location method is applied to automatically extract events from detailed, publicly available outage data of investor-owned utilities in Massachusetts \cite{MassachusettsData}.
Thresholds of $t_{max}=3$ hours and $d=10$ km are used.
Data from NOAA's storm events database \cite{noaaStormsDB}, DOE's OE-417 form, and local weather reports are used to verify whether the automatically extracted events correspond to actual weather events. 
Table~\ref{tab:eventsData} shows that the top 10 largest events extracted automatically from the outage data correspond closely to major weather events.
Notably, three distinct March 2018 Nor'easters appear in the top ten events. These are correctly separated by the algorithm because sufficient time elapsed between successive storms for all outages from the earlier storm to be restored before the next storm began, breaking the temporal overlap chain. This serves as useful validation: the algorithm did not over-aggregate these closely spaced storms into a single multi-week event, which time-only grouping with large $t_{max}$ would risk.

\begin{table}[hbpt]
\caption{Weather events associated with automatically extracted events from outage data} 
\label{tab:eventsData}
\centering
\begin{tabular}{@{}c c c l@{}}
Total & Start Date and Time & Duration & Weather Event \\
Outages     & (ET)                  & (hours)    & Details \\ \hline
7511        & 2021-10-26 15:29   & 179.1      & Oct. 2021 Nor'easter \\
6863        & 2018-03-02 04:35   & 177.9      & March 2018 Nor'easter \\
4783        & 2018-03-13 02:28   & 127.6      & March 2018 Nor'easter \\
4515        & 2018-03-07 18:39   & 180.7      & March 2018 Nor'easter \\
2894        & 2017-10-29 20:13   & 113.7      & Oct. 2017 Nor'easter \\
2594        & 2020-08-04 13:00   & 129.3      & Hurricane Isaias \\
2527        & 2013-02-08 14:48   & 149.1      & Winter Storm Nemo \\
2435        & 2019-10-16 22:23   & 91.7       & Coastal Storm \\
1923        & 2020-04-13 09:51   & 67.2       & Easter Tornado\\
1564        & 2016-02-05 07:13   & 67.9       & Winter Storm \\
\end{tabular}
\end{table}

\begin{figure}[htb]
    \centering
    \includegraphics[width=1.0\linewidth]{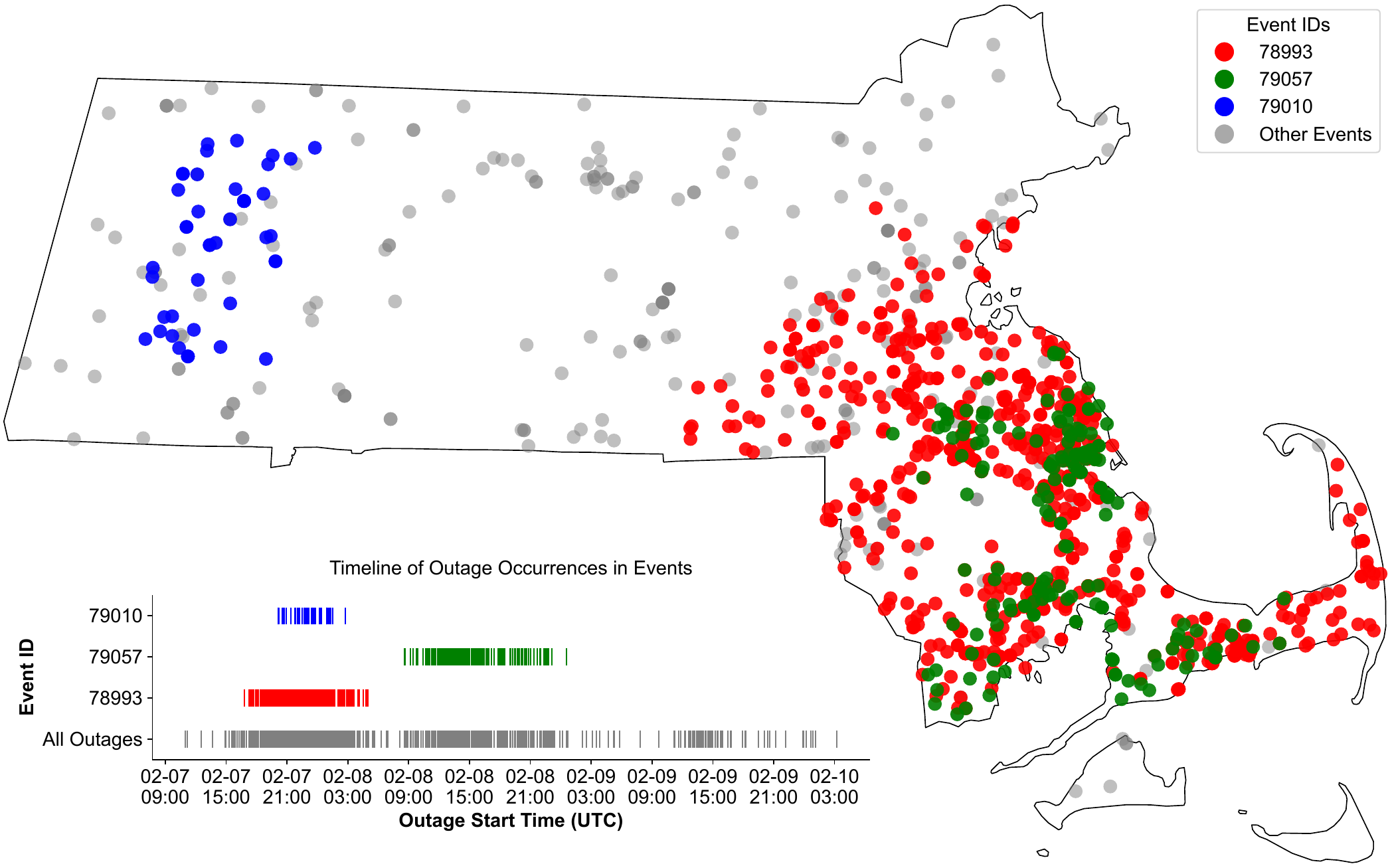}
    \caption{Example of events extracted by time and location grouping using detailed outage data. Only the top three largest events (by the number of outages) are shown. The inset plot shows the timeline of outages in the top three largest events.}
    \label{fig:exampleEventDetailed}
\end{figure}

Fig.~\ref{fig:exampleEventDetailed} shows an example of an actual event extracted using the time-and-location-based grouping method from the Massachusetts detailed outage data. 
The 1266 outages shown are grouped into a single event when only time-based grouping (section \ref{sec:eventstime}) is used. However, when both time and location are used, the outages are grouped into 174 distinct events, the top three of which are shown in Fig.~\ref{fig:exampleEventDetailed}. The red and green points show outages that occur in close proximity to each other, but the inset timeline plot shows they do not overlap in time, which is why the algorithm separated them into two events. Similarly, the red and blue bars in the timeline overlap in time; however, when we look at their locations, we notice they are not close to each other, which is why the algorithm did not group them into the same event. The gray points indicate outages that are grouped into smaller, mostly single-outage events.

\subsection{EAGLE-I outage data results}
\begin{figure}[htb]
    \centering
    \includegraphics[width=1.0\linewidth]{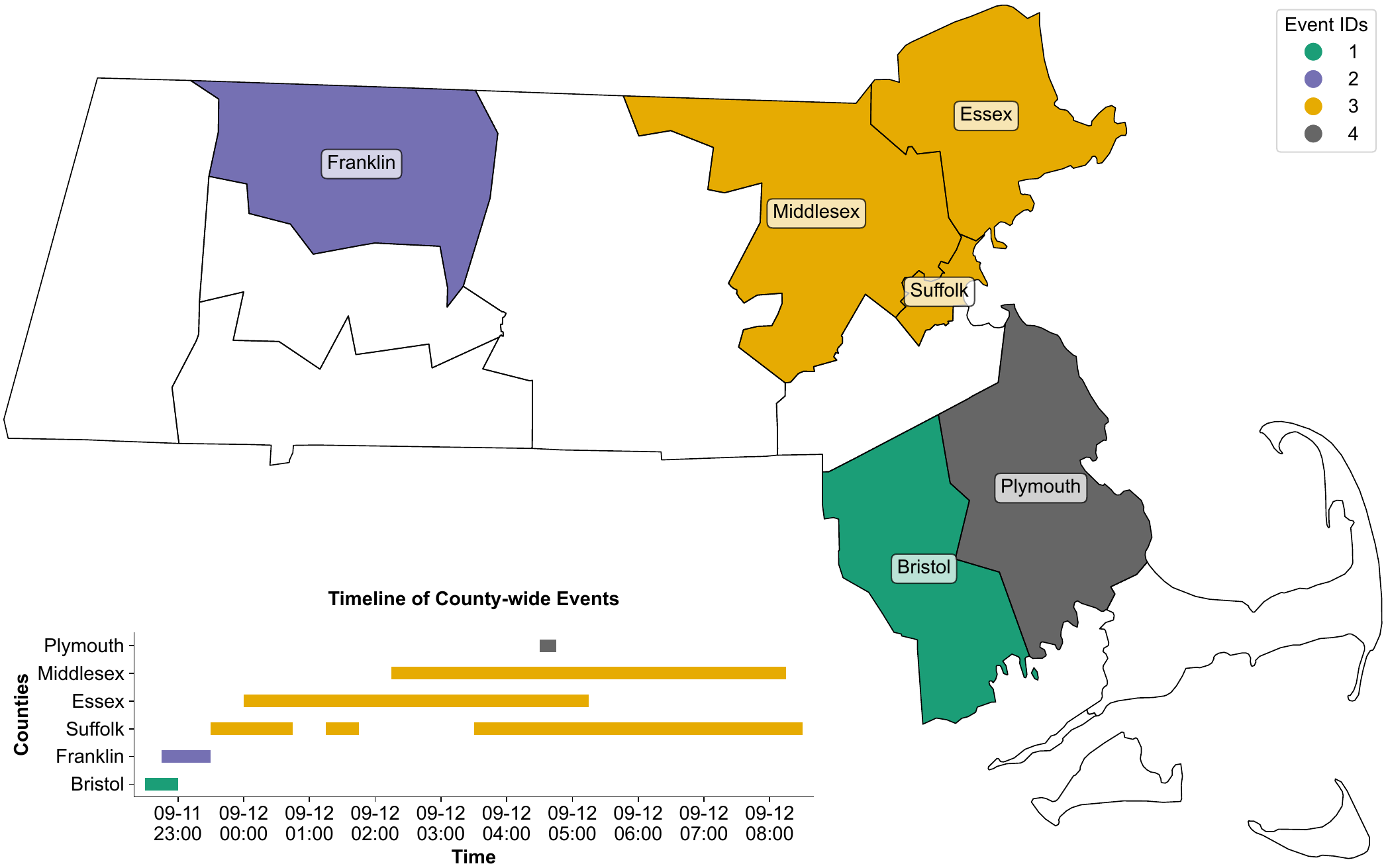}
    \caption{Example of events extracted by time-and-location-based grouping using EAGLE-I outage data. The inset plot shows the timeline and duration of each of the 8 county events.}
    \label{fig:exampleEventEAGLEi}
\end{figure}

Fig.~\ref{fig:exampleEventEAGLEi} shows an example of applying the proposed time-and-location-based grouping method to the EAGLE-I data from the state of Massachusetts.
Since the data is aggregated at the county level, the outages within a county can be grouped into events using the time-only grouping method. However, when conducting a statewide or multi-county analysis, it is imperative to use the time-and-location grouping method to ensure events are correctly grouped. 
Fig.~\ref{fig:exampleEventEAGLEi} shows 8 county events that overlap in time\footnote{The timeline plot shows that there are 3 county events in Suffolk county.} and would be grouped together into a single event if only time-based grouping is used. However, when we use time-and-location-based grouping, these 8 county events are combined into 4 events.
We can compare the location of county events and their timelines shown in the inset plot in Fig.~\ref{fig:exampleEventEAGLEi} to illustrate the time-and-location-based grouping logic: the county event in Franklin overlaps in time with the county event in Bristol, but they are not in neighboring counties, which is why they are considered separate events.

\begin{figure}[htb]
    \centering
    \includegraphics[width=1.0\linewidth]{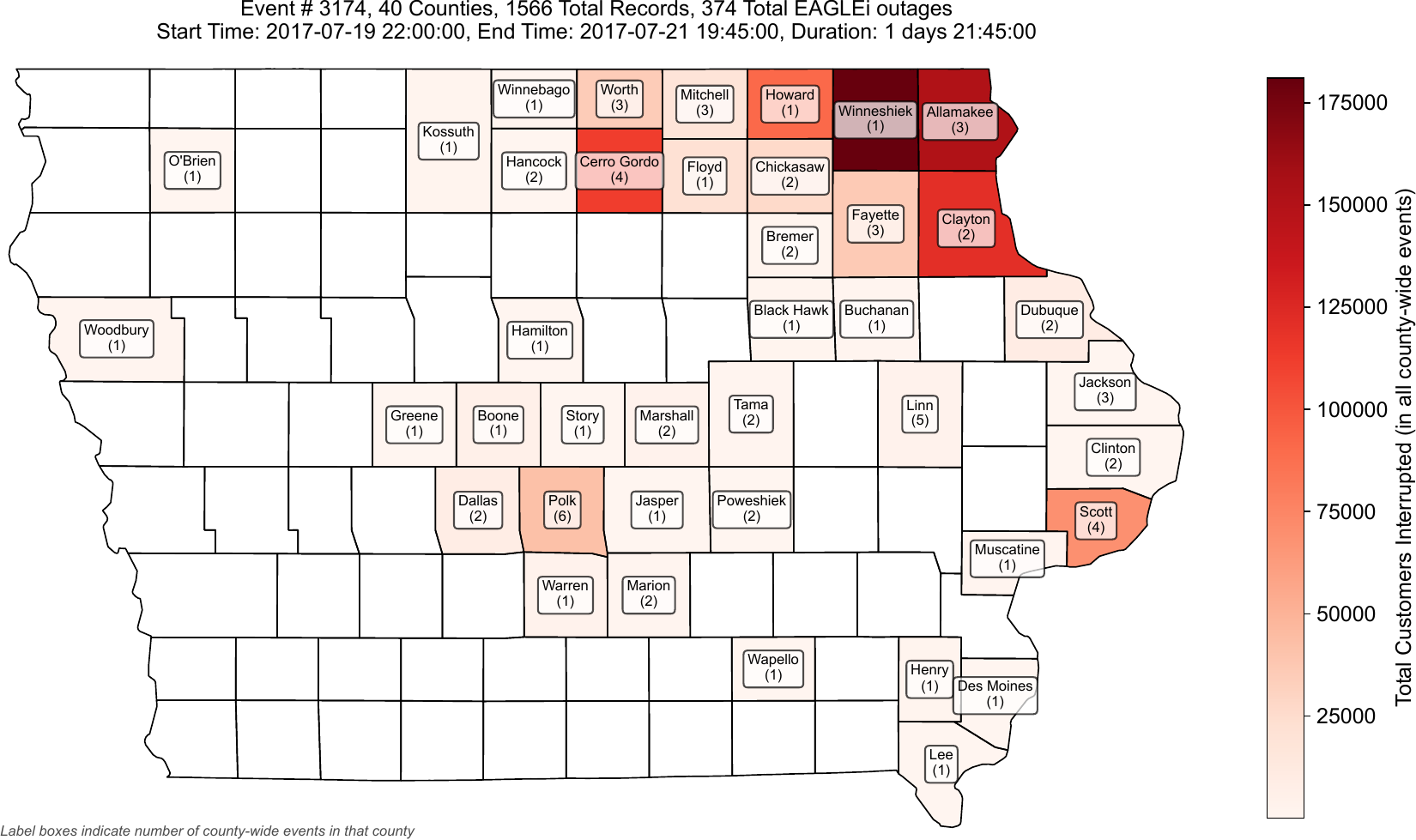}
    \caption{Example of an event from EAGLE-I data from Iowa. The event contains 76 county events, grouped into a single event using time-based grouping. The numbers within brackets under each county's name represent the total number of county events in that county.}
    \label{fig:2ndexampleEventEAGLEi}
\end{figure}

\begin{figure}[htb]
    \centering
    \includegraphics[width=1.0\linewidth]{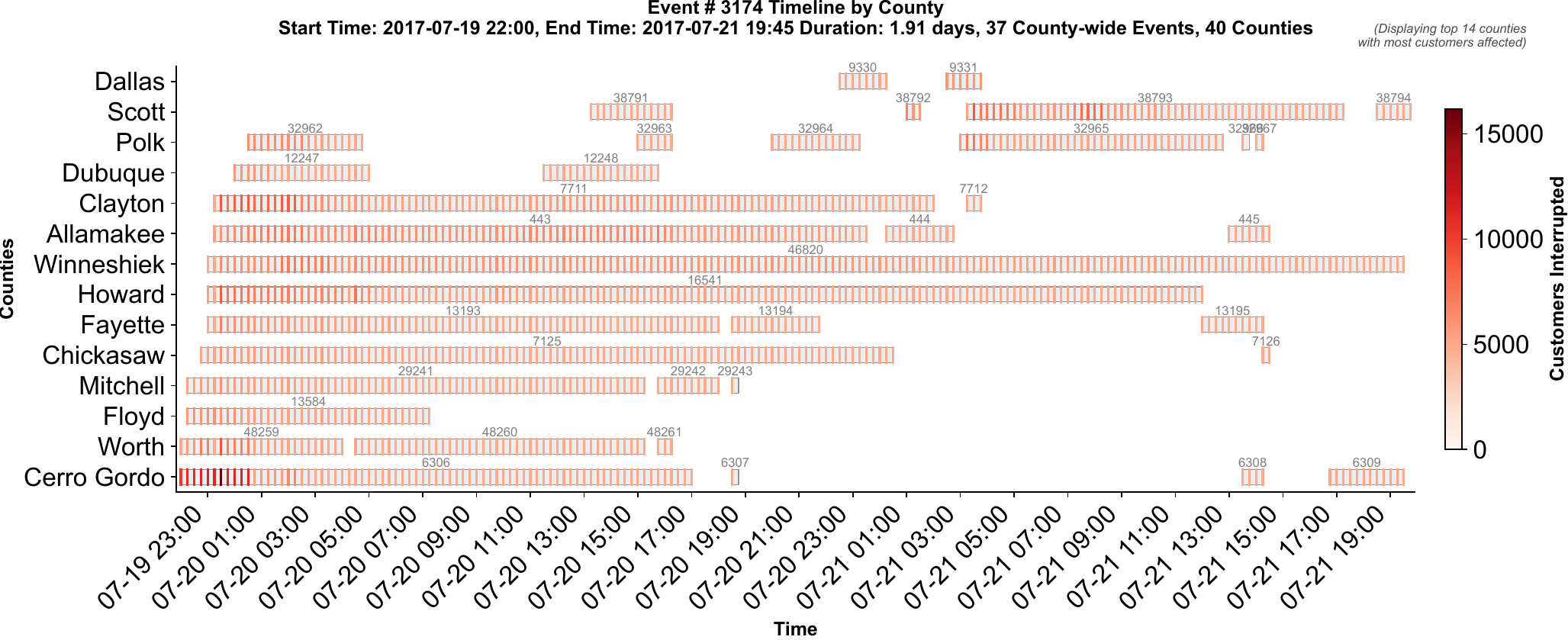}
    \caption{Timeline plot of the county events involved in the event shown in Fig.~\ref{fig:2ndexampleEventEAGLEi}. Although the event involves 40 counties, only the 14 with the largest number of customers affected are shown in this plot due to space limitations. Each horizontal bar corresponds to a 15-minute timestamp, with the bar's color intensity indicating the number of interrupted customers at that timestamp. Consecutive horizontal bars represent a county event, which is created using the time-based grouping for the EAGLE-I data, discussed in Section~\ref{sec:eventsFromEagleiData}.}
    \label{fig:2ndexampleEventEAGLEiTimeline}
\end{figure}

\begin{figure}[htb]
    \centering
    \includegraphics[width=1.0\linewidth]{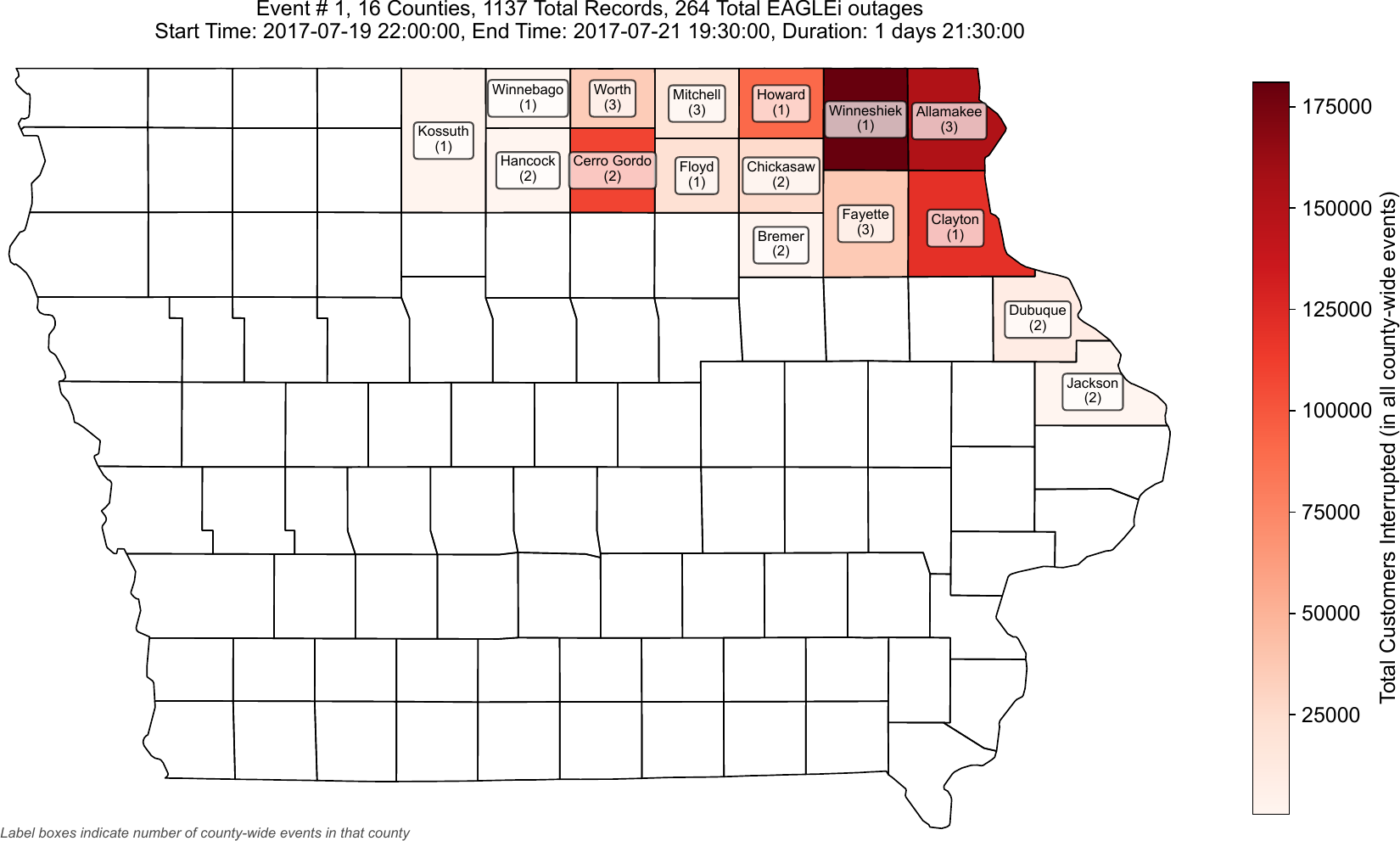}
    \caption{Event created by applying the time-and-location-based grouping to the 76 county events shown in Fig.~\ref{fig:2ndexampleEventEAGLEi}. The time-and-location-based grouping kept only 30 county events in this event and grouped the others into smaller events (not shown in this plot) because they did not satisfy the time-overlap and location-closeness conditions.}
    \label{fig:2ndexampleEventEAGLEiSpatiotemporal}
\end{figure}

\begin{figure}[htb]
    \centering
    \includegraphics[width=1.0\linewidth]{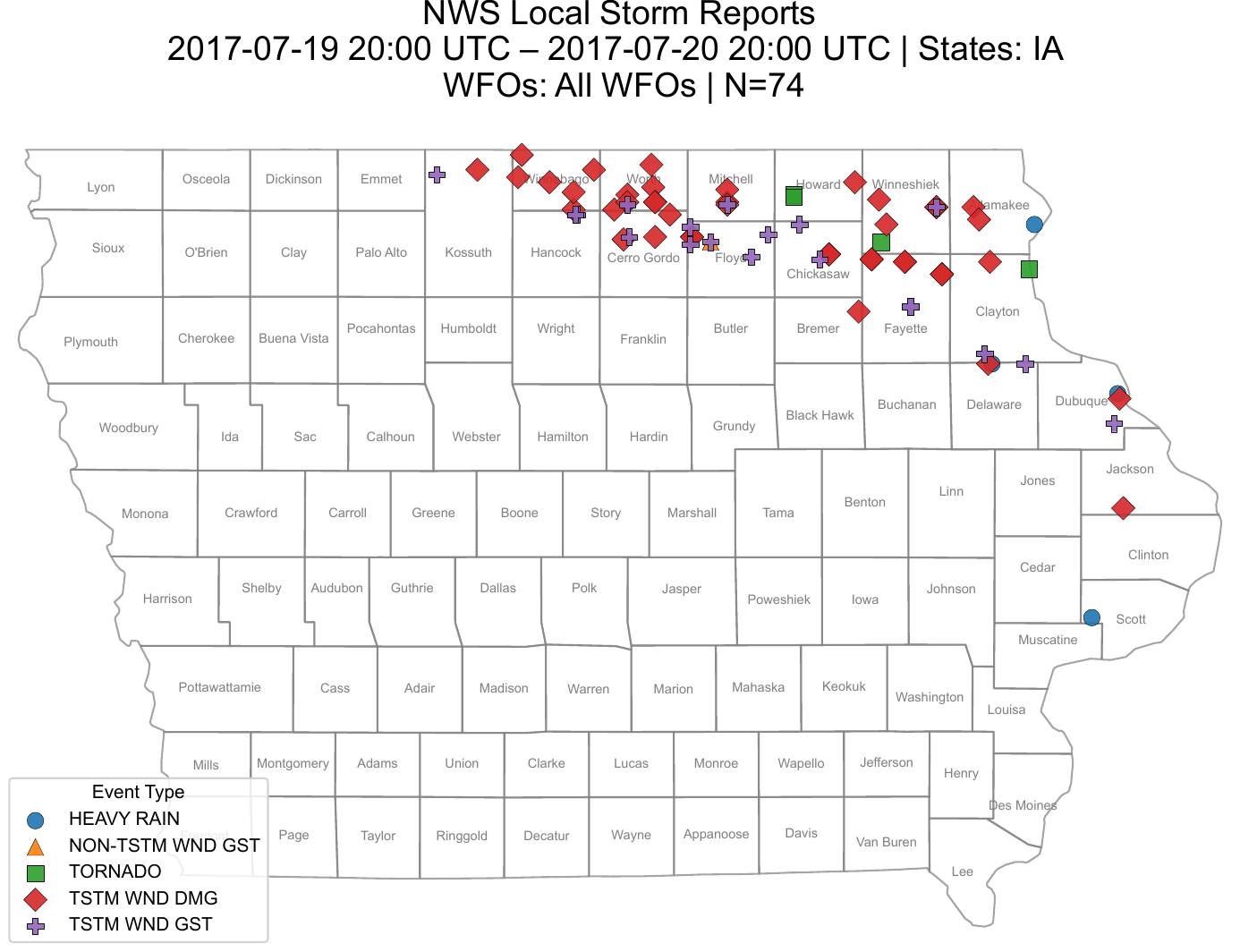}
    \caption{Storms reported during the time window of the event shown in Fig.~\ref{fig:2ndexampleEventEAGLEiSpatiotemporal}. A large number of thunderstorms with damaging winds and a few tornadoes were reported in north-eastern Iowa during this time.}
    \label{fig:2ndexampleEventEAGLEiWeather}
\end{figure}

Fig.~\ref{fig:2ndexampleEventEAGLEi} shows another example of an event extracted from the EAGLE-I data for the state of Iowa. 
The time-based method groups together 
76 county events that occurred in 40 counties. 
Fig.~\ref{fig:2ndexampleEventEAGLEiTimeline} also shows the corresponding timeline of county events in the event (only the top 14 counties with the largest number of customers affected are shown in the plot to limit space). The timeline plot verifies that all the involved county events overlap in time and hence are correctly grouped. However, when we look at the location map of the county events in Fig.~\ref{fig:2ndexampleEventEAGLEi}, we see that many of the county events occurred in counties quite far from each other. 
Therefore, when we apply the time-and-location-based grouping to the same set of 76 county events, these are split into multiple events, the largest of which is shown in Fig.~\ref{fig:2ndexampleEventEAGLEiSpatiotemporal}. 
This largest event consists of 30 county events which satisfy the time and location thresholds of the time-and-location-based grouping, whereas the rest of the 46 county events are grouped by the algorithm into multiple smaller, geographically separate events concentrated in other parts of the state (not shown in Fig.~\ref{fig:2ndexampleEventEAGLEiSpatiotemporal}).
To validate that the event shown in Fig.~\ref{fig:2ndexampleEventEAGLEiSpatiotemporal} represents the correct grouping of related outages, we look at the weather during the event time window. Fig. \ref{fig:2ndexampleEventEAGLEiWeather} shows the storms reported during the event time window, sourced from the NOAA Local Storm Reports (LSR) \cite{LsrIEM}. We see that there was bad weather only in the north-eastern part of the state during the respective time window, so that the time-and-location-based grouping method based only on outage data correctly grouped outages that were related to the bad weather.


\begin{figure}[htb]
    \centering
    \includegraphics[width=1.0\linewidth]{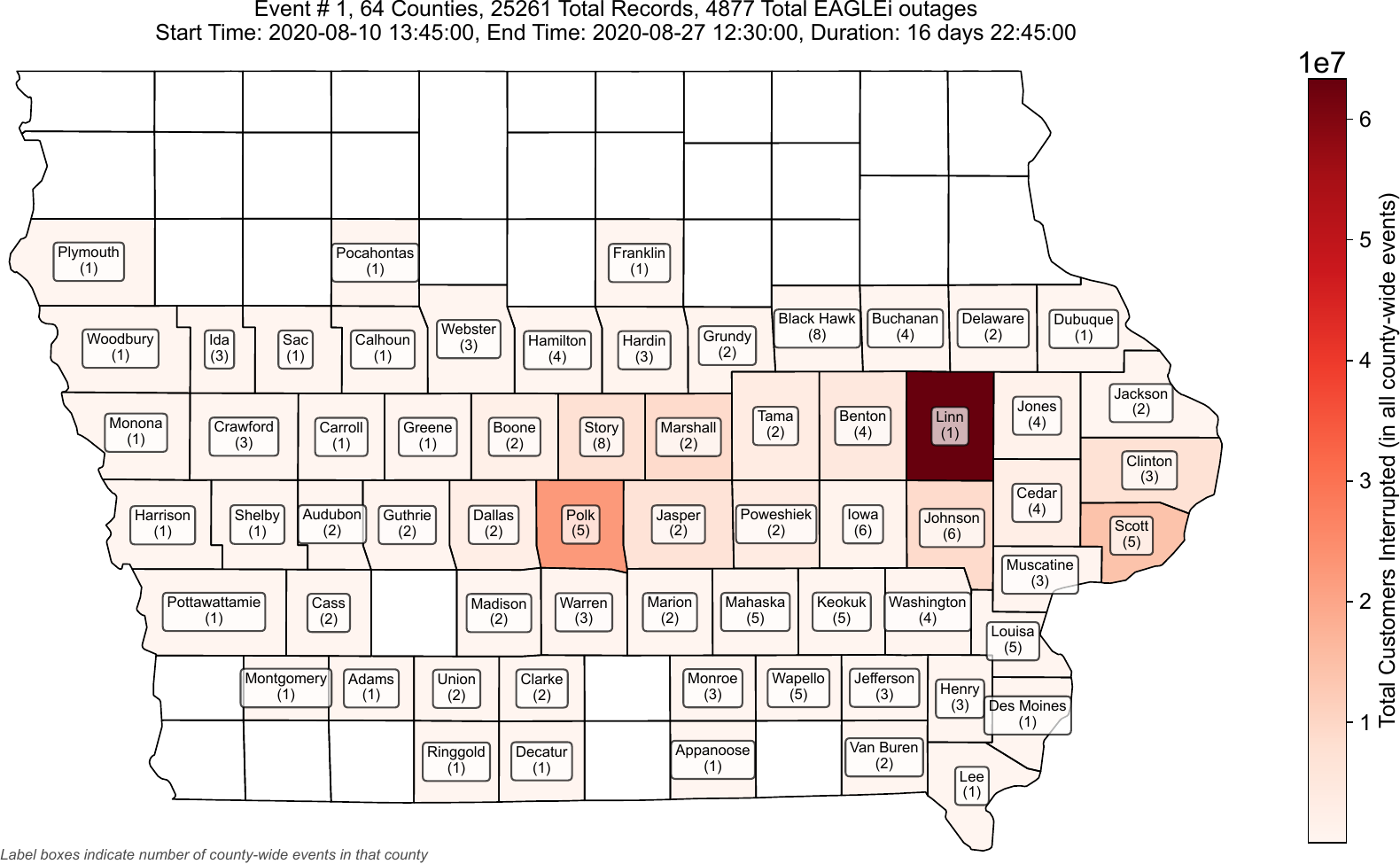}
    \caption{All the county events associated with the August 2020 Midwest Derecho are automatically grouped into one event by the time-and-location-based grouping applied to the EAGLE-I data.}
    \label{fig:derechoEAGLEi}
\end{figure}

\begin{figure}[htb]
    \centering
    \includegraphics[width=1.0\linewidth]{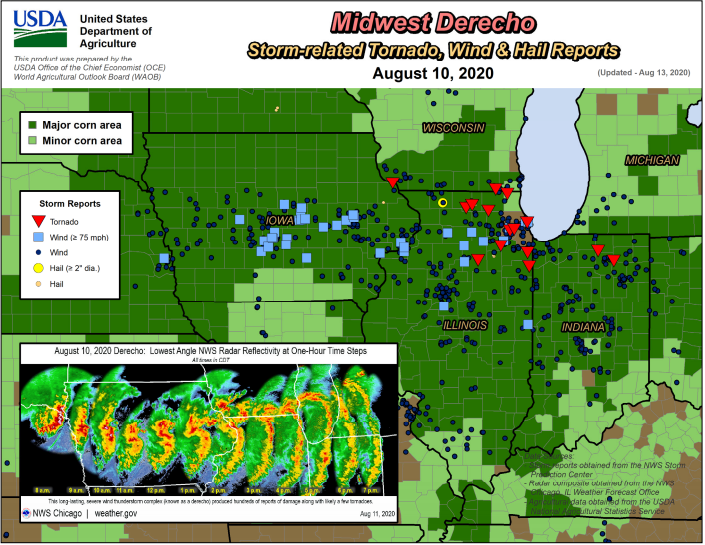}
    \caption{Tornadoes, and wind \& hail reports during the August 2020 Midwest derecho (source: USDA \cite{derechoUSDA}).}
    \label{fig:derechoEAGLEiWeather}
\end{figure}

Another example of an event extracted using the time-and-location-based grouping from the EAGLE-I data for the state of Iowa is shown in Fig.~\ref{fig:derechoEAGLEi} and corresponds to the August 2020 Midwest Derecho, one of the largest events to occur recently in the Midwest. Fig.~\ref{fig:derechoEAGLEiWeather} shows the tornadoes, and wind and hail reports during the derecho. A comparison of Figs.~\ref{fig:derechoEAGLEi} and \ref{fig:derechoEAGLEiWeather} shows that the time-and-location-based grouping nicely grouped all the county events in Iowa that are associated with the derecho into a single event.


\section{Conclusion}\label{sec:conclusion}

We explain an overlapping outage principle used to group outages into events based on time, and generalize it to group outages into events based on both time and location. The main contributions are:
\begin{itemize}
    \item Events are formed by a unifying principle of overlaps of outages in time and location that can be visualized as overlapping cylinders,
    \item Applying the method to detailed utility outage data as well as  web-scraped EAGLE-I data,
    \item Demonstrating the value of also using location information when analyzing outages in wide geographic areas,
    \item Analyzing the time and distance thresholds used in forming the events and developing a method and metric to tune the thresholds to get credible events.
\end{itemize}
Each outage in detailed outage data is represented as a cylinder in three-dimensional space, with a disk of radius $d/2$ in the geographic plane and a bounded vertical extent corresponding to the limited outage duration $[o,\min\{r,o+t_{\rm max}\}]$. 
Two outages that are close in both time and location are exactly those that have overlapping cylinders. 
Outages are grouped into the same event if they are linked by a series of overlapping cylinders. 
This leads directly to a graph-theoretic formulation in which outages are nodes, cylinder overlaps are edges joining outages, and resilience events are the connected components of the resulting graph. This formulation is precise and computationally tractable, and it admits standard optimizations such as interval trees, sweep-line algorithms, k-d trees, and grid hashing for large-scale datasets.

The events in web-scraped EAGLE-I outage data are formed with similar principles of overlap in time and location, but work with county events in which performance curves for outages in the county exceed a threshold. County events overlap in time if their outage processes overlap, and overlap in location if the counties are neighbors.

Grouping outages into resilience events is crucial for resilience analysis of utility outage data. When grouping is performed by time alone over a large area, outages far from a given storm that are either isolated outages or from a different storm can be merged into a single event. This overestimates event size, geographic span, and duration, and it introduces bias into downstream analyses such as fragility curve estimation, restoration rate modeling, and infrastructure investment prioritization. The proposed method, combining timing and location, produces more credible groupings of outages into events better correlated with storm data and therefore forms events that are a more reliable basis for quantitative resilience assessment.

Any systematic resilience analysis of a large number of outage records requires automatic event grouping, as it is impractical to manually group outages into events beyond a limited number of selected cases. Any automatic or manual scheme of grouping outages into events makes judgments about marginal cases. And any automatic grouping applied to all cases will make a few judgments that some experts could disagree with. 
Therefore it can be appropriate to have expert review of a practically limited number of the automatically processed events, while acknowledging the impracticality of systematic manual processing, and the efficiency of checking the automatically processed events (almost all of which will be within the range of expert judgments) rather than manually processing from scratch.
We conclude that automatic processing is clearly needed and that, while perfection is not attainable, it is clearly important that automatic processing perform very well.  Improvements to automatic processing are worthwhile.

The proposed approach to systematically and automatically extract credible events from utility data provides an improved foundation for the quantitative study of widespread resilience events driven by real data. 
In particular, it suppresses anomalous events, removes unrelated outages far away from weather events, and extends analyses of detailed outage data and \hbox{EAGLE-I} data to widespread events beyond the county level.

\printbibliography

@ARTICLE{AbdelmalakIACCESS23,
  author={Abdelmalak, Michael and Cox, Jordan and Ericson, Sean and Hotchkiss, Eliza and Benidris, Mohammed},
  journal={IEEE Access},  
  title={Quantitative Resilience-Based Assessment Framework Using {EAGLE-I} Power Outage Data}, 
  year={2023},
  volume={11},
  pages={7682-7697},
  doi={10.1109/ACCESS.2023.3235615}
}

@INPROCEEDINGS{AbdelmalakRW22,
  author={Abdelmalak, Michael and Ericson, Sean and Cox, Jordan and Ben-Idris, Mohammed and Hotchkiss, Eliza},
  booktitle={Resilience Week}, 
  title={A Power Outage Data Informed Resilience Assessment Framework}, 
  year={2022},
  pages={1-6},
  doi={10.1109/RWS55399.2022.9984016}}

@ARTICLE{LeeIACCESS24,
  author={Lee, Sangkeun Matthew and Chinthavali, Supriya and Bhusal, Narayan and Stenvig, Nils and Tabassum, Anika and Kuruganti, Teja},
  journal={IEEE Access}, 
  title={Quantifying the Power System Resilience of the {US} Power Grid Through Weather and Power Outage Data Mapping}, 
  year={2024},
  volume={12},
  pages={5237-5255},
  doi={10.1109/ACCESS.2023.3347129}}

@article{DunnNHR18,
author = {Dunn et al., Sarah},
title = {Fragility curves for assessing the resilience of electricity networks constructed from an extensive fault database},
journal = {Natural Hazards Rev.},
volume = {19},
%number = {1},
pages = {04017019},
year = {2018},
doi = {10.1061/(ASCE)NH.1527-6996.0000267},
URL = {https://ascelibrary.org/doi/abs/10.1061/%28ASCE%29NH.1527-6996.0000267}
}

@article{PapicAS20,
author = {Papic, Milorad and Ekisheva, Svetlana and Cotilla-Sanchez, Eduardo},
title = {A risk-based approach to assess the operational resilience of transmission grids},
journal = {Applied Sciences},
volume = {10},
number = {14},
pages = {4761},
year = {2020}
}

@inproceedings{EkishevaPMAPS22,
author = {Ekisheva, Svetlana and I. Dobson and J. Norris and R. Rieder },
title = {Assessing transmission resilience during extreme weather with outage and restore processes},
booktitle = {Probalistic Methods Applied to Power Systems},
month={06},
location={Manchester UK},
year={2022}
}

@ARTICLE{CarringtonPS21,
  author={Carrington, Nichelle’Le K. and Dobson, Ian and Wang, Zhaoyu},
  journal={IEEE Trans. Power Systems}, 
  title={Extracting Resilience Metrics From Distribution Utility Data Using Outage and Restore Process Statistics}, 
  year={2021},
  volume={36},
  number={6},
  pages={5814-5823},
  doi={10.1109/TPWRS.2021.3074898}}

@INPROCEEDINGS{DonaldsonCIRED26,
  author={Donaldson, Daniel L and Ahmad, Arslan and Dobson, Ian},
  booktitle={CIRED Workshop on Implementing Successful Innovation in Distribution Networks}, 
  title={Comparing the extreme weather event definitions
used to quantify power system resilience}, 
  year={2026},
  volume={},
  number={},
  pages={},
  doi={},
  address={Brussels, Belgium}
}

@misc{MassachusettsData,
    author = {{Commonwealth of Massachusetts}},
    title = {Power outages},
    howpublished = {Available: \url{https://www.mass.gov/info-details/power-outages}},
    url = {https://www.mass.gov/info-details/power-outages},
    note = {Accessed: Feb. 10, 2025},
%  year = {2025}
}

@misc{eagleiWebsite, 
    author={{DOE}},
    title = {{EAGLE-I}},
    howpublished = {Available: \url{https://eagle-i.doe.gov/}},
    url={https://eagle-i.doe.gov/},
    publisher={U.S. Department of Energy},
    note = {Accessed: May. 12, 2026}
}

@article{henryRESS12,
title = {Generic metrics and quantitative approaches for system resilience as a function of time},
journal = {Reliability Engineering \& System Safety},
volume = {99},
pages = {114-122},
year = {2012},
issn = {0951-8320},
doi = {https://doi.org/10.1016/j.ress.2011.09.002},
url = {https://www.sciencedirect.com/science/article/pii/S0951832011001748},
author = {Devanandham Henry and Jose {Emmanuel Ramirez-Marquez}}
}

@article{DobsonPS23,
  author={Dobson, Ian},
  journal={IEEE Transactions on Power Systems}, 
  title={Models, Metrics, and Their Formulas for Typical Electric Power System Resilience Events}, 
  year={2023},
  volume={38},
  number={6},
  pages={5949-5952},
  doi={10.1109/TPWRS.2023.3300125}}

@techreport{keenNREL24,
  title={Current practices in distribution utility resilience planning for wildfires},
  author={Keen, Jeremy and Matsuda-Dunn, Reiko and Krishnamoorthy, Gayathri and Clapper, Haley and Perkins, Lila and Leddy, Laura and Grue, Nick},
  year={2024},
  institution={National Laboratory of the Rockies, Golden, CO USA}
}

@techreport{keenNREL24b,
  title={Current Practices in Distribution Utility Resilience Planning for Winter Storms},
  author={Jeremy Keen and Reiko Matsuda-Dunn and Gayathri Krishnamoorthy and Haley Clapper and Lila Perkins and Laura Leddy and Nick Grue},
  year={2024},
  institution={National Laboratory of the Rockies, Golden, CO USA},
  doi = "10.2172/2478837",
}

@techreport{keenNREL24c,
  title={Current Practices in Distribution Utility Resilience Planning for Hurricanes and Non-Winter Storms},
  author={Jeremy Keen and Reiko Matsuda-Dunn and Gayathri Krishnamoorthy and Haley Clapper and Lila Perkins and Laura Leddy and Nick Grue},
  year={2024},
  institution={National Laboratory of the Rockies, Golden, CO USA},
  doi = "10.2172/2478839",
}

@article{IEEEstd1782,
  author={},
  journal={{IEEE} Std 1782-2022 (Revision of {IEEE} Std 1782-2014)}, 
  title={{IEEE} Guide for Collecting, Categorizing, and Utilizing Information Related to Electric Power Distribution Interruption Events}, 
  year={2022},
  volume={},
  number={},
  pages={1-113},
  doi={10.1109/IEEESTD.2022.9882080}}

@INPROCEEDINGS{MorrisPMAPS16,
  author={Morris, Euan A. and Bell, Keith R. W. and Elders, Ian M.},
  booktitle={Probabilistic Methods Applied to Power Systems}, 
  title={Spatial and temporal clustering of fault events on the {GB} transmission network}, 
  year={2016},
  volume={},
  number={},
  pages={1-9},
  doi={10.1109/PMAPS.2016.7764087}}

@article{WardCC13,
  title={The effect of weather on grid systems and the reliability of electricity supply},
  author={Ward, David M},
  journal={Climatic Change},
  volume={121},
  number={1},
  pages={103--113},
  year={2013},
  publisher={Springer}
}

@article{wangSR26,
  title={Data-driven quantification and visualization of resilience metrics of power distribution systems},
  author={Wang, Dingwei and Maharjan, Salish and Zheng, Junyuan and Liu, Liming and Wang, Zhaoyu},
  journal={Scientific Reports},
  year={2026},
  publisher={Nature Publishing Group UK London}
}

@misc{noaaStormsDB,
  author = {{National Oceanic and Atmospheric Administration (NOAA)}},
  title = {{Storm Events Database | National Centers for Environmental Information}},
  url = {https://www.ncei.noaa.gov/stormevents/},
  howpublished = {Available: \url{https://www.ncei.noaa.gov/stormevents/}},
  note = {Accessed: Mar. 24, 2026}
}

@ARTICLE{PanteliProcIEEE17,
  author={Panteli, Mathaios and Trakas, Dimitris N. and Mancarella, Pierluigi and Hatziargyriou, Nikos D.},
  journal={Proc. IEEE}, 
  title={Power Systems Resilience Assessment: Hardening and Smart Operational Enhancement Strategies}, 
  year={2017},
  volume={105},
  number={7},
  pages={1202-1213},
  doi={10.1109/JPROC.2017.2691357}
}

@book{Pantelibook26,
  title={Fundamentals of Power System Resilience: Disruptions by Natural Causes},
  author={Panteli, Mathaios and Moreno, Rodrigo and Trakas, Dimitris and Jamieson, Magnus and Mancarella, Pierluigi and Strbac, Goran and Hatziargyriou, Nikos},
  year={2026},
  publisher={John Wiley \& Sons}
}

@article{StankovicPS23,
  title        = {Methods for analysis and quantification of power system resilience},
  author       = {A. Stankovic~et~al.},
  year         = {2023},
  journal      = {IEEE Trans. Power Systems},
  number       = {5},
  pages        = {4774--4787},
  issue        = {38}
}

@article{NanRESS17,
title = {A quantitative method for assessing resilience of interdependent infrastructures},
journal = {Reliability Engineering \& System Safety},
volume = {157},
pages = {35-53},
year = {2017},
issn = {0951-8320},
doi = {https://doi.org/10.1016/j.ress.2016.08.013},
url = {https://www.sciencedirect.com/science/article/pii/S095183201630374X},
author = {Cen Nan and Giovanni Sansavini}
}

@techreport{nercSOR23,
    author = {NERC},
    title = {2023 {S}tate of {R}eliability Technical Assessment, {J}une 2023},
    institution = {North American Electric Reliability Corp.},
    url = {https://www.nerc.com/pa/RAPA/PA/Performance%20Analysis%20DL/NERC_SOR_2023_Technical_Assessment.pdf},
}

@misc{LsrIEM,
  author       = {{Iowa State University}},
  title        = {{Iowa Environmental Mesonet: Archived Local Storm Reports}},
  howpublished = {\url{https://mesonet.agron.iastate.edu/request/gis/lsrs.phtml}},
}

@misc{derechoUSDA,
  author       = {{USDA}},
  title        = {{USDA} Weekly Weather and Crop Bulletin - Vol. 107 No. 33 - {A}ugust 18, 2020},
  year = {2020},
  howpublished = {Accessed 06/04/2026. Available online at \url{https://downloads.usda.library.cornell.edu/usda-esmis/files/cj82k728n/3t946c39r/w6634s119/wwcb3320.pdf}},
}

@article{leeIEEEaccess25,
  title={A Near-Real-Time Model for Predicting Electricity Disruptions in Texas During Winter Storms},
  author={Lee, Jangjae and Lee, Sangkeun and Chinthavali, Supriya and Paal, Stephanie},
  journal={IEEE Access},
  year={2025},
  publisher={IEEE}
}

@article{ShuaiEST25,
  title={Electric Power Reliability, Energy Burdens, and Climate Change Beliefs in the United States},
  author={Shuai, Hang and Chen, Chien-Fei and Sovacool, Benjamin and Sumkhuu, Suzanna and Shen, Zhenglai},
  journal={Environmental Science \& Technology},
  year={2025},
  publisher={ACS Publications}
}

@article{RahmanERC25,
  title={Characteristics of power outages from compound weather extremes in Florida},
  author={Rahman, Mohammad Siddiqur and Wahl, Thomas and Nagaraj, Meghana and Enr{\'\i}quez, Alejandra R},
  journal={Environmental Research: Climate},
  volume={4},
  number={3},
  pages={035008},
  year={2025},
  publisher={IOP Publishing}
}

@article{StanishevskaArxiv25,
  title={Early-Warning of Thunderstorm-Driven Power Outages with a Two-Stage Machine Learning Model},
  author={Stanishevska, Iryna},
  journal={arXiv preprint arXiv:2510.03959},
  year={2025}
}

@misc{doe417,
  author       = {{U.S. Department of Energy}},
  title        = {{DOE-417 Electric Emergency Incident and Disturbance Report}},
  howpublished = {Available: \url{https://doe417.pnnl.gov/}},
  url          = {\url{https://doe417.pnnl.gov/}},
  note         = {Accessed: 2026-06-11}
}

@techreport{climatecentral22,
  author      = {{Climate Central}},
  title       = {{Surging Power Outages and Climate Change}},
  institution = {Climate Central},
  year        = {2022},
  month       = {9},
  url         = {https://www.climatecentral.org/report/surging-power-outages-and-climate-change},
  note        = {Accessed: 2026-06-11}
}

@article{AnkitSSR22,
title = {{U.S.} Resilience to large-scale power outages in 2002–2019},
journal = {Journal of Safety Science and Resilience},
volume = {3},
number = {2},
pages = {128-135},
year = {2022},
issn = {2666-4496},
doi = {https://doi.org/10.1016/j.jnlssr.2022.02.002},
url = {https://www.sciencedirect.com/science/article/pii/S2666449622000147},
author = {Aman Ankit and Zhanlin Liu and Scott B. Miles and Youngjun Choe}
}

@article{LarsenENERGY16,
title = {Recent trends in power system reliability and implications for evaluating future investments in resiliency},
journal = {Energy},
volume = {117},
pages = {29-46},
year = {2016},
issn = {0360-5442},
doi = {https://doi.org/10.1016/j.energy.2016.10.063},
url = {https://www.sciencedirect.com/science/article/pii/S0360544216314979},
author = {Peter H. Larsen and Kristina H. LaCommare and Joseph H. Eto and James L. Sweeney}
}

@article{AhmadPS24,
  title        = {Towards using utility data to quantify how investments would have increased the wind resilience of distribution systems},
  author       = {Ahmad, Arslan and Dobson, Ian},
  year         = {2024},
  journal      = {IEEE Trans. Power Systems},
  volume       = {39},
  number       = {4},
  pages        = {5956--5968},
  doi          = {https://doi.org/10.1109/TPWRS.2023.3342729}
}

@article{ahmadArxiv26a,
      title={Quantifying resilience for distribution system customers with {SALEDI}}, 
      author={Arslan Ahmad and Ian Dobson},
      year={2026},
      eprint={xxxx},
      journal={arXiv preprint},
      primaryClass={xxxx},
      url={https://arxiv.org/abs/2602.07684},
      note={ https://arxiv.org/pdf/2602.07684}
}

@ARTICLE{DobsonPS24,
  author={Dobson, Ian and Ekisheva, Svetlana},
  journal={IEEE Trans. Power Systems}, 
  title={How Long is a Resilience Event in a Transmission System?: Metrics and Models Driven by Utility Data}, 
  year={2024},
  volume={39},
  number={2},
  pages={2814-2826},
  doi={10.1109/TPWRS.2023.3292328}}

\end{document}